# Analyzing the basic principles of tissue microarray data measuring the cooperative phenomena of marker proteins in invasive breast cancer

Horst Buerger[1]
Florian Boecker[2]
Jens Packeisen[3]
Konstantin Agelopoulos[4]
Kathrin Poos[2]
Walter Nadler[5]
Eberhard Korsching[2]

[1]Institute of Pathology, Paderborn, Germany; [2]Institute of Bioinformatics, University of Münster, Münster, Germany; [3]Institute of Pathology, Osnabrück, Germany; [4]Department of Medicine, Hematology and Oncology, University of Münster, Münster, Germany; [5]Institute for Advanced Simulation (IAS), Jülich Supercomputing Centre (JSC), Jülich, Germany

**Background:** The analysis of a protein-expression pattern from tissue microarray (TMA) data will not immediately give an answer on synergistic or antagonistic effects between the observed proteins. But contrary to apparent first impression, it is possible to reveal those cooperative phenomena from TMA data. The data is (1) preserving a lot of the original physiological information content and (2) because of minor variances between the tumor samples, contains several related slightly different biological states. We present here a largely assumption-free combinatorial analysis, related to correlation networks but with much less arbitrary constraints. A strong focus was put on the analysis of the basic data to analyze how the cooperative phenomena might be imprinted in the TMA data structure.

**Results:** The study design was based on two independent panels of 589 and 366 invasive breast cancer cases from different institutions, assembled on tissue microarrays. The combinatorial analysis generates an optimal rank ordering of protein-expression coherence. The outcome of the analysis corresponds to all the single observations scattered over several publications and integrates them in one context. This means all these scattered observations can also be deduced from one TMA experiment. A comprehensive statistical meta-analysis of the TMA data suggests the existence of a superposition of three basic coherence situations, and offers the opportunity to analyze these data properties with additional real-world data and synthetic data in more detail.

**Conclusion:** The presented algorithm gives molecular pathologists a tool to extract dependency information from TMA data. Beyond this practical benefit, some light was shed on how dependency aspects might be imprinted into expression data. This will certainly foster the refinement of algorithms to reconstruct dependency networks. The implementation of the algorithm is at the moment not end-user suitable, but available on request.

**Keywords:** tissue microarrays, protein expression, dependency structure, breast cancer, progression, algorithm, biological networks

## Introduction

The cellular system consists of a large number and a wide variety of interacting molecules. At a certain moment, the presence and quantity or absence of molecules defines a certain state of the cell. This, for a certain state-typical expression pattern of molecules, is defined by a vast and in many aspects poorly understood network of interactions between these molecules.[1] Types of molecules might be macromolecules like RNA, DNA, proteins, and also fragments thereof.

A broad spectrum of approaches has recently been proposed to unravel the dependency structure between selected types of macromolecules.[2] All those approaches are based in a certain way on expression data: means relative concentrations or

Correspondence: Eberhard Korsching
Institute of Bioinformatics,
University Hospital of Münster,
12–14 Niels-Stensen-Strasse,
Münster 48149, Germany
Tel +49 25 1835 0651
Fax +49 25 1835 3005
Email korschi@uni-muenster.de

derivatives thereof.[3] The analysis strategy may focus on more abstract aspects, like pure coherence of a certain (maybe itself abstract) class of events, eg, integrative networks,[4] or on modeling dependency by similarity in expression patterns, eg, by correlations.[5] Many approaches will be refined by filtering for indirect effects, eg, partial correlations (basics and application),[3,6] and/or introduce more or less justified additional constraints like thresholds and additional downstream procedures.[7] There are of course a lot more approaches, like modeling concrete biochemical reactions by differential equations,[8] but we focus here on the fundamental aspects of (correlation-based) coexpression systems.

Most of the approaches are based on massive parallel measurements of molecular expression states, but few to no (biological) replicates. We center on the opposite, a limited number of proteins but with many biological replicates (several hundred). The measurements are based on immunohistochemical data from tissue microarrays (TMAs), a data source that has not been used for advanced dependency studies up to now.

In the time of next-generation sequencing approaches, TMAs seem to be old-fashioned, but that is obviously not true. In contrast to gene-expression microarrays and next-generation sequencing approaches, TMAs contain a lot more relevant information of the biological context. Looking on the sample source of procedures to measure expression rates, it can be noticed that they are based mostly on tissue extracts. TMAs in contrast preserve the tissues and therefore expose additional spatial and morphological information.[9–11] Moreover, TMAs measure proteins, which define more reliably the physiological (steady state) situation of a cell type in a preserved tissue. At present, this technique is used to define and apply characteristic patterns of marker proteins for diagnostic purposes.[12] Some other studies, showing the power of TMAs in the determination of prognostic and predictive markers in breast cancer, are reviewed in Chatterjee et al[13] and Packeisen et al.[11] However, so far most studies have not considered the potential of TMAs to characterize more complex (de)regulation mechanisms in cancer tissues. Instead, they were considered only as an approach complementary to expression microarrays.[14,15] This situation inspired us to analyze the basic properties of TMA data in more depth.

The evaluation of the TMA is not only based on the measured protein signal strength but also on additional information like localization (which cellular compartment or area of the extracellular matrix is stained), morphology (tissue/cellular architecture), and cellular neighborhood (which cell types are present at all). So the TMA technique provides a number of factors that add more specificity to the analyzed (patho)physiological situation. This will implicitly reduce the combinatorial complexity of presumed protein dependencies. With TMAs, we are analyzing more dependencies that truly exist in the experimental situation we are interested in, and avoid facing the task of filtering out a lot of the hypothetical but not implemented dependencies of the analyzed biological system. Of course, there are also drawbacks. The training of the pathologist doing the evaluations has a major impact on the quality of the results. The sensitivity of the TMAs might be limited for certain proteins, which will result in an underestimation of those proteins.

So, how are we utilizing existing concepts in the analysis of TMA data? The methodological approaches in expression-data analysis use at the first step real-scale data and the Pearson correlation, but very early switch to binary or nominal scale data, so losing quality and information content. The correlation step is more or less used as a feasible proximity measure, but not further questioned concerning the process properties. We perform here similar to Langfelder et al.[16] Due to the complexity of the resulting correlation space, and the need to get the correlation values ordered, this approach uses additional threshold or filtering procedures and/or is simplifying the scale of the data (a "binary" concept). But this simplification appears to be highly questionable. Most of the expression values populate a very narrow expression range. Differences might be found not by discriminating between signal strength, but by finding unique compositions of factors making up a certain expression module. So we switch at that point to a more data-driven approach. We employ a combinatorial optimization procedure, and try to find the best configuration of dependencies concerning two selected subgroups of factors. The concept allows intermediate states, and the model is not limited to a certain type of network structure or biological data.

To give the analysis some statistical importance, we composed two large independent collections of invasive breast cancer cases from the Gerhard Domagk Institute of Pathology, University of Münster (589 cases: collection 1), and the Institute of Pathology, Osnabrück (366 cases: collection 2), containing invasive breast cancer cases graded from G1 to G3. All these cases were assembled on TMAs. The biological background of this analysis was the diagnostic/prognostic importance of cytokeratin (CK) expression patterns in physiological breast cells and invasive breast cancer cells, as can be seen in a multitude of publications throughout the last few years, eg, Abd

El-Rehim et al[17] and Korsching et al.[18] The measurements of the CKs were complemented by other factors known to play or supposed to play a role in this invasive cancer scenario.

Our objective was to (1) uncover the cooperativeness of the expression profiles of protein markers in a (patho) physiological state of invasive breast cancer, (2) to show characteristics of the used TMA data giving clues to extract systematically more information from TMA data by conventional means, (3) to support the reported tracer function of the CK expression in breast cancer progression, and (4) to analyze the properties of the measured data structure in more detail, to understand how cooperativeness is imprinted in the replicated expression data.

# Methods

## Ethics statement

Data were analyzed anonymously. Nonetheless, we conducted this study according to the principles expressed in the Declaration of Helsinki. The study was approved by the Institutional Review Board of the University of Münster. We acquired tissue samples only with the informed consent of the patients or patients' next of kin, with the understanding by all parties that it may well be used for research. All patients provided written informed consent for the collection of samples and subsequent analysis.

## Tissue samples

Two independent series of invasive breast cancer cases graded from G1 to G3 were used for this study. The clinicopathological features of the patient samples can be seen in Table 1. Paraffin-embedded tissue of 589 invasive breast cancer cases, originating from the archives of the Gerhard Domagk Institute of Pathology, University of Münster (collection 1) and 366 breast cancer cases originating from the archives of the Institute of Pathology, Osnabrück (collection 2) were used for the production of a TMA in accordance with published protocols (Kononen et al,[19] Packeisen et al[20]). Each carcinoma was represented by two cores. The spot diameter was 0.6 mm. The distance between the spots was 1.0 mm. A specialized tissue array precision instrument (Beecher Instruments, Silver Spring, MD, USA) was used to assemble the TMA. Sections of 3 μm thickness were cut from the TMA blocks and stained. The average value of both cores was taken for further investigation. To enable the definition of representative tumor areas, a hematoxylin and eosin-stained section was made from each donor block. The tumor series represented all stages, grades, and histological subtypes of invasive breast cancer in accordance with the literature (data not shown here).

Staining procedures were done according to standard protocols. The pretreatment conditions, the source, and the dilution of the commercially available primary antibodies are listed in Table S1. The immunohistochemistry results were evaluated independently by two pathologists according to well-established scoring schemes.[21]

## Data preprocessing

A certain number of measurements (7% for the first and 17% for the second collection, randomly distributed) missed the stringent evaluation scheme, or were lost during the staining procedure. The missing values in the data matrix were replaced with the median value of the remaining immunohistochemical values of that type. Some impact on the variance of the results can be estimated from Figure S2A (variance of the solid symbols). But presumably a much greater impact comes from the collectors and pathologists themselves. The choice of the median or the mean for missing values did not influence the final order of the test factors (details not presented).

## Optimization algorithm

The goal of the presented algorithm (master plan, see Figure 1, column A) is to construct an optimal rank order of a group of test factors for multiple reference situations. The rationale behind this is to stabilize the results by reducing the number of possibilities to generate such a result. This is comparable to a scientist who is making a couple of different experiments and not only one to extract knowledge about the system of interest.

The algorithmic approach was developed based on two independent invasive breast cancer collections (Table 2). The data groups are further subdivided by the composition of measured and included molecular factors. This diversity was used first and foremost to evaluate the stability of the results. But some other considerations, like the impact of inserting molecular factors or assembling several dependency clusters, are also fostered by this experimental design. For validation purposes, it was important for us that some of the expression levels of the chosen factors and their regulatory dependencies were already described in the literature.[22,17]

Apart from the permutation parts, the algorithm is implemented on the basis of the open-source statistical software R (www.r-project.org) and a Linux platform. The time-critical parts were implemented in Fortran (Fortran Compiler version 11 for Linux. Intel, Santa Clara, CA, USA).

We start with a classical array representation of the data where the columns show the immunohistochemical variables

**Table 1** Clinicopathological features of the patient samples

| Tumor grade | T-category | N-category |
|---|---|---|
| All (n = 955) | | |
| G1 = 163 | T1a = 12 | N negative = 544 |
| G2 = 475 | T1b = 63 | N positive = 411 |
| G3 = 317 | T1c = 390 | |
| | T2 = 374 | |
| | T3 = 65 | |
| | T4 = 51 | |
| Münster (n = 589) | | |
| G1 = 111 | T1a = 21 | N negative = 312 |
| G2 = 291 | T1b = 35 | N positive = 277 |
| G3 = 187 | T1c = 247 | |
| | T2 = 227 | |
| | T3 = 36 | |
| | T4 = 23 | |
| Osnabrück (n = 366) | | |
| G1 = 65 | T1a = 7 | N negative = 197 |
| G2 = 175 | T1b = 25 | N positive = 169 |
| G3 = 126 | T1c = 150 | |
| | T2 = 142 | |
| | T3 = 25 | |
| | T4 = 17 | |

(factors) and the rows represent the analyzed individual samples (Figure 1, 1A). The factors are divided into a reference set and a test set (Figure 1, 2A). For the particular case where the CKs should be analyzed, the reference factors were filled with CK5, CK8/18, CK14, CK19, CK1, and CK10. Some of these CKs are known to discriminate subentities of invasive breast carcinomas. All the other factors belong to the test-factor group.

In the next step, a proximity measure is calculated between the reference and test candidates. Here we used the correlation according to Pearson.[23] The reason for that choice is our interest in differentiating the long and complex immunohistochemical profiles of the measured biological factors (see Figure S1 for details of the effect).

The correlation cross-tabulation (Figure 1, 3A) with the n reference factors annotated on top and the m test factors annotated to the side is the basis of the following permutation procedure (Figure 1, 4A). The aim is to rank the m test factors in the n reference situations in a way that the global sum of the sum of squares ($ssq_g$) of n linear regression approaches (one for each reference situation) will be minimized (Figure 2). The constraint is that the rank ordering of the m test factors is identical for all n situations and a certain permutation. The resulting best situation will describe the best observable dependency of the test factors to the ensemble of reference factors.

So in our case, we are not primarily interested in a single correlation value, but in an ensemble of correlation values and their optimal rank ordering under the constraint of an optimal fit to multiple targets (reference factors). Figure S2, C and D shows why the number of reference factors makes sense. The distance to a random situation will increase.

## Tests of statistical significance

The statistical significance of our results was tested using resampling methods. The general procedure of such tests is that a superset of surrogate data sets is generated. The test statistic in question is then obtained for each of those new data sets, and the value of the test statistic of the original data set is checked against its distribution in the superset of surrogate data sets.

The distribution of the test statistic over the surrogate data superset corresponds to the null hypothesis that – given the constraints of the experiment – the observed value is due only to random fluctuations.

A test-statistic value for the original data set lying far in the tail of that distribution is a sign of high significance, which can also be quantified using its position in the distribution. The *P*-value obtained corresponds to the fraction of entries in that histogram that is still further from the bulk than the test-statistic value obtained from the original data set.

Surrogate data sets were generated using both the permutation and the bootstrap methods.[24,25] In permutation tests, the surrogate data sets are generated by randomly shuffling the complete original data set, thereby reproducing exactly the statistical distribution of the original data. In bootstrap testing, surrogate data sets are generated by randomly choosing data (with replacement) from the original data set, separately for each immunohistochemical factor. Here, the surrogate data sets reproduce the statistical distribution of the original data only on average. The idea behind the latter approach is that the original data set is already an heuristic approximation of the underlying population the experiment is performed on. Thereby, fluctuations due to the stochastic nature of an experiment, ie, random observations from the underlying population, are accounted for, too.

Usually, we used a superset of $10^{+4}$ surrogate data sets, each of the original data set size, and we used both, permutation and bootstrap methods, in each case. The test statistic we computed in each case was the minimal $ssq_g$ for a data

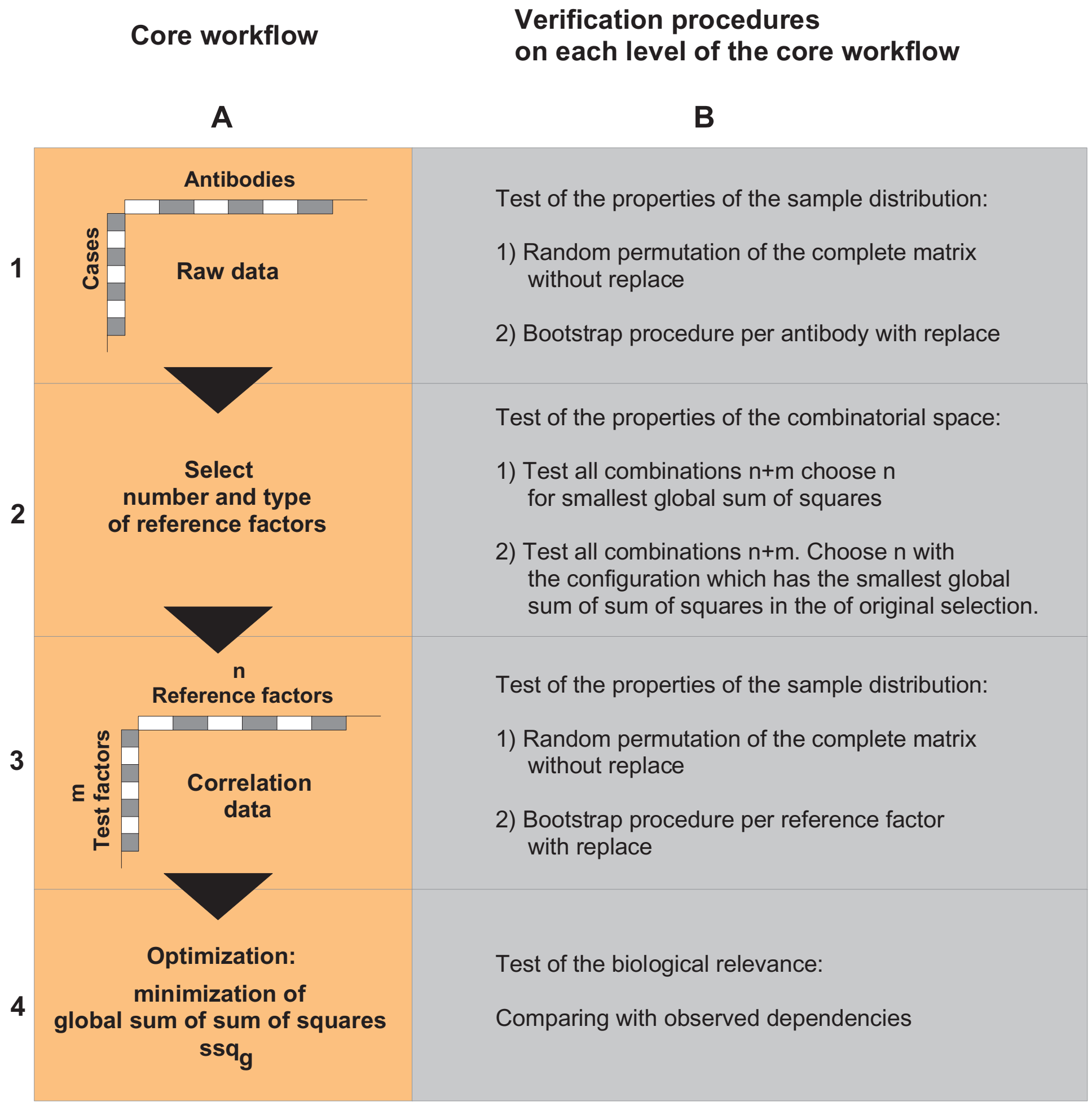

**Figure 1** Block diagram of the algorithm and of the verification procedures.

**Notes:** This overview summarizes on the left the important core steps of the algorithm. The column on the right summarizes all the verification procedures as well as the additional analytic procedures. **Column A:** this flowchart explains the processing of the data from the initial raw-data matrix until the end of the permutation procedure. The main analysis is based on an a priory chosen set of biological factors – the cytokeratins. After raw-data processing, the data is collected in a two-dimensional matrix – patient cases versus factors (**1A**). The factors are subdivided in two groups: a test set and a reference set (**2A**). In the next step, the correlation is calculated for all combinations of the test and reference set (**3A**). Then, for each of the possible permutations of test factors, the global sum of the sum of squares ($ssq_g$) for all the analyzed situations is calculated (**4A**). The situation with the minimal $ssq_g$ is used to choose the optimal rank order. **Column B:** here, the different verification procedures of the algorithm are presented. In (**1B** and **3B**), two verifications will be performed on the data matrix: a permutation sampling approach of the complete raw-data or correlation matrix, and a bootstrap approach on factor columns of the raw-data matrix or reference-factor columns of the correlation matrix. In (**2B**), the spotlight is on all the possible group compositions of the size six of 16 giving insight into all the data properties, and in (**4B**) the biological relevance of the results will be discussed in relation to published data.

set, obtained from this data set's optimal rank ordering, as described above.

## Distribution of the minimal $ssq_g$ of different group compositions

We started the analysis with a selection of six CKs acting as reference factors. The remaining 10 factors were used as test factors. The CKs were selected as reference factors, because we already own independent a priory knowledge on their biological role. This opens up the opportunity to verify the behavior of the algorithm. But there are many other six of 16 partitions remaining. To get some insight into the whole dependency structure we analyzed also all these remaining six of 16 partitions. Every unique reference combination will result, as described above, in an optimal $ssq_g$. The distribution of all best $ssq_g$ will show the best six of 16 composition. This distribution was also analyzed according to the mode structure and stability.

The basic six of 16 approach was generalized to examine alternate numbers of reference factors on the same raw data set. The experiments ranged from reference group sizes of eight of 16 to four of 16. The more extreme values below 4 were not feasible due to computing time constraints. Limited insight into the computing times is given in supplementary Figure S3.

**Table 2** Data groups

| Acronym | Groups | Number of cases | Number of factors | Composition |
|---|---|---|---|---|
| (M-16) | Münster | 589 | 16 | A and B |
| (O-16) | Osnabrück | 366 | 16 | A and C |
| (M-13) | Münster | 589 | 13 | A |
| (O-13) | Osnabrück | 366 | 13 | A |
| (M + O-13) | Münster and Osnabrück | 955 | 13 | A |

**Notes:** Two different sample collections were used. The first set is based on 589 cases of the pathology of Münster and the second on 366 cases of the pathology of Osnabrück, in total, 955 cases. With these two collections, several molecular markers were measured. **A:** CK1, CK10, CK5, CK14, CK8/18, CK19, vimentin, epidermal growth factor receptor, MIB1, erb-B2, p53, estrogen receptor, progesterone receptor; **B:** BCL2, cyclin D1, epithelial membrane antigen; **C:** smooth-muscle actin, p63, CD10. In the last column of the table, the composition of factors for each data group is indicated. The acronyms will be used throughout the publication.

It has to be noted that the present limit is about 18–20 factors, assuming all permutations should be calculated.

# Results

## Definition of used raw data sets

The protein expression of 16 factors was examined, in the cells of interest, in invasive breast cancer cases from two pathologies (collection 1 and collection 2). The measurements resulted in 16 relative protein quantifications for each sample. Three of the antibodies differed between the two collectives; therefore, five basic situations were analyzed as depicted in Table 2.

## Optimal rank ordering based on 589 invasive breast cancer cases

Will the rank order of collection (M-16) reflect already-published measurements of other research groups? The reference set comprises CK1, CK5, CK8/18, CK10, CK14, and CK19, known to be involved in breast cancer progression pathways. The test set was built by a couple of more or less prominent markers like the epidermal growth factor receptor (EGFR), vimentin, proliferation-related factor Ki-67 (antibody name MIB1), tumor suppressor p53, oncogene c-erb-B2, interacting mediator of cell death BCL2, cyclin D1, epithelial membrane antigen, progesterone receptor (PR), and estrogen receptor (ER).

The result of the optimization process is shown in Figure 3A1. The panel with the six plots, each illustrating one of the reference situations, represents the examined data in the state with the minimal $ssq_g$ error. The resulting optimal rank order on the $x$-axis reflects the dependency of the test factors on the reference factors. In contrast to a pure ranking of correlation coefficients, the positions of several test factors were adjusted according to the fit for multiple reference situations. The final and high order for all factors can be seen in Figure 3A2.

Even though the individual correlation values might be rather low, a clear differential effect could be seen for CK5, CK14, CK19, and CK8/18, while the correlation values for CK1 and CK10 were rather low, and the respective regression lines did not show a clear tendency. These two CKs do not contribute to the differential behavior. Additionally, we can observe that the center positions of the optimal rank order of the test factors on the $x$-axis also do not contribute to the differential behavior. Only the extreme parts define the synergistic or antagonistic dependency.

## Verification of the optimal rank ordering with an independent invasive breast cancer data collection

The verification was performed with collection O-16 (Table 2). Thirteen of these antibodies are identical to those used in collection 1. The reference set comprises again CK1, CK5, CK8/18, CK10, CK14, and CK19. The test set consists again of EGFR, vimentin, Ki-67 (MIB1), p53, erb-B2, PR, ER, and three different factors: CD 10, transcriptional modulator p63, and smooth-muscle actin.

Three objectives were pursued: (1) how well does a smaller and independent invasive breast cancer sample collective from a different pathology prove the results of the main collective, (2) to analyze whether and how different factors influence the result, and (3) how do the combination of the concordant measurements of the invasive breast cancer sample collectives behave. The optimal rank order of the independent breast cancer data collection is shown in Figure 3B1 and 2. The same basic principle as in Figure 3A1 from the first breast cancer data collection can be observed. There is a minor rearrangement of the sequence of the test factors on the $x$-axis due to the three different proteins in the test set. CK14 exhibits a weaker response. The higher $ssq_g$ value compared to Figure 3A2 might also reflect a higher heterogeneity of the sample collective and/or number of missing values. This is also apparent in Figure 3B2, where the order is not as clearly defined as in Figure 3A2. Nevertheless, the type of order is very similar to the first breast cancer data collection, and the strong effectors concerning the synergistic or antagonistic behavior remain the same.

To analyze the effect of the three different factors in more detail, we made for both breast cancer collections a design where the test set is reduced to the seven factors present in both collectives (Figure S4C and D, based on collections

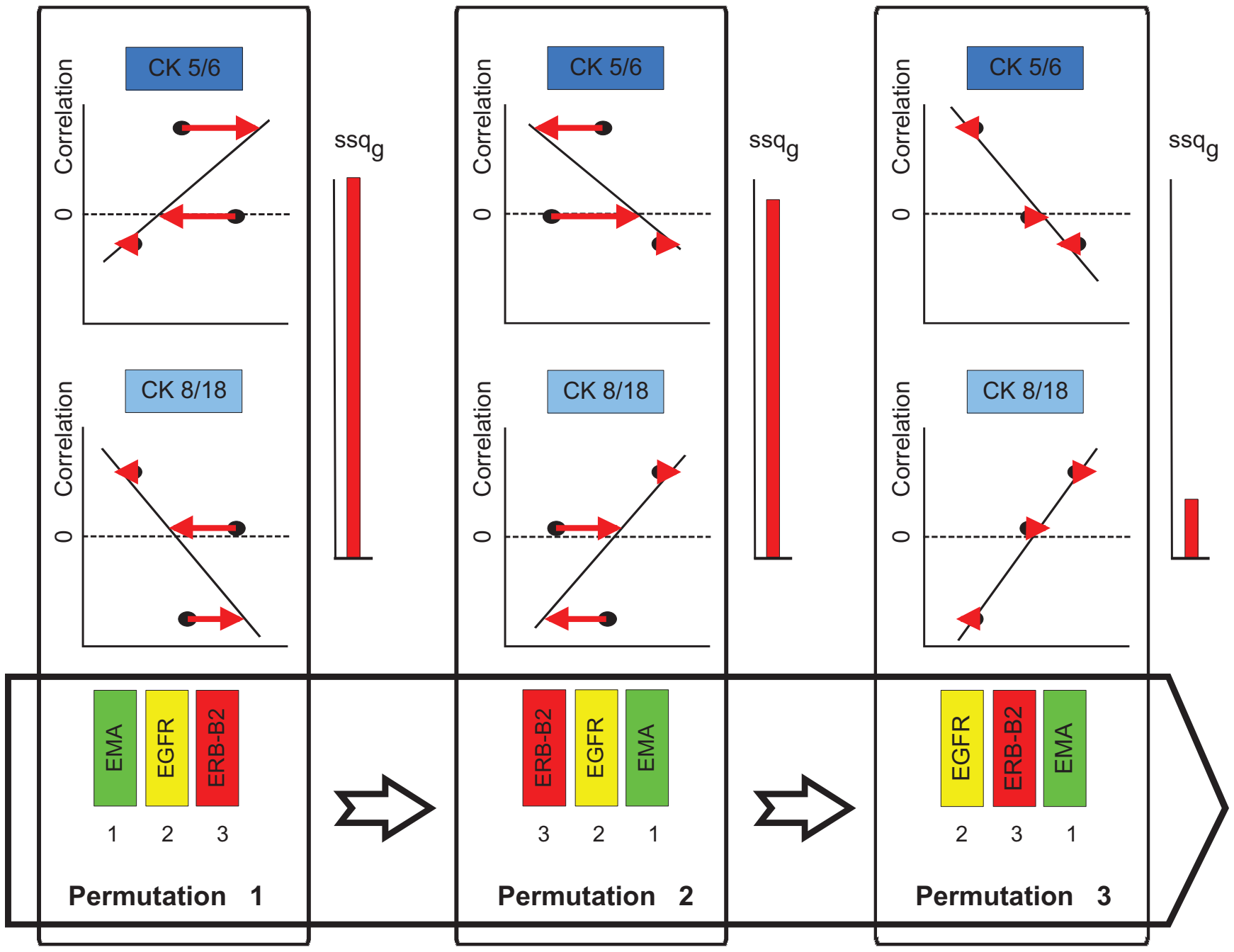

**Figure 2** Details of the optimization process and the optimal rank order.
**Notes:** This time-critical procedure was implemented with the Fortran Compiler version 11 for Linux (Intel, Santa Clara, CA, USA). Some of the Fortran routines are from the open-source Fortran library of John Burkardt. Here, a simplified version of the permutation procedure is presented. Two reference factors building two reference situations (symbolized by dark- and light-blue boxes) and three test factors (green, yellow, red boxes) are shown in three idealized permutation cycles. At each step, the global sum of the sum of squares ($ssq_g$) gets smaller, ending up at a minimum (indicated to the right of each permutation step). On the *x*-axis, the permutations are shown. We start with permutation 1 (n = 2: cytokeratin [CK] 5/6 and CK8/18). The start point of the optimization process has the given order of the m test factors. For that rank ordering, a linear regression curve is calculated for each reference factor resulting in two sums of squares. The global sum of those sums ($ssq_g$) is the decision value for the ongoing permutation procedure and should be optimized towards the lowest possible value. This is done in this example by choosing a new permutation of test factors. A new $ssq_g$ is calculated (see permutation 2). If this new result is smaller than the previous result, this new result will be stored together with the corresponding rank order. This process continues until all permutations (or the unique 50%) are tested (see permutation 3: idealized final and best state in our example). As "optimal rank ordering," we consider the permutation that gives the smallest $ssq_g$. The list of all permutations of the test factors is obtained by complete enumeration.[34]
**Abbreviations:** EGFR, epidermal growth factor receptor; EMA, epithelial membrane antigen.

M-13 and O-13). The omission of three different factors from the test-set results in either case in a deletion in the center part of the optimal rank order, but does not alter the remainder. The effects on the overall structure are very small.

Joining both breast cancer data collections on the basis of the corresponding 13 factors (M + O-13) resulted in a similar outcome, as observed in the breast cancer collection M-13 (Figure S4E).

Despite all minor differences, which account to variations in the sample cohort, different numbers of missing values, and three different proteins, the analysis led to the same results in both invasive breast cancer collectives.

## Data sets and their distributions

In this and the following sections, we analyze the structure and properties of the data sets more basically, to understand the principles behind how dependencies might model the given data structure.

First, we look at the distributions of the five basic analysis scenarios depicted in Table 2. In Figure 4, the primary $ssq_g$ distributions for collections M-16 and O-16 are presented. All possible orders contribute with one $ssq_g$ to these distributions. Results for M-13, O-13, and M + O-13 can be seen in Figure S5. Depending on the number of $ssq_g$ values, the smoothness of the distributions varies as well as the median. All five distributions of the respective $ssq_g$ values share the same characteristic of a left-skewed distribution. The shape and height of the distribution is slightly different between the two sample collections. Also in these cases, an impact of the borderline cases, handled differently in both pathologies, and the effects from the three different proteins must be taken into consideration.

## Analyzing the differences between the optimal rank ordering of the original data and the random and bootstrap samples of these data

These approaches should test whether or not the results are generated by chance. The optimization algorithm calculated a minimal $ssq_g$ for each scenario of Table 2. Disturbing the

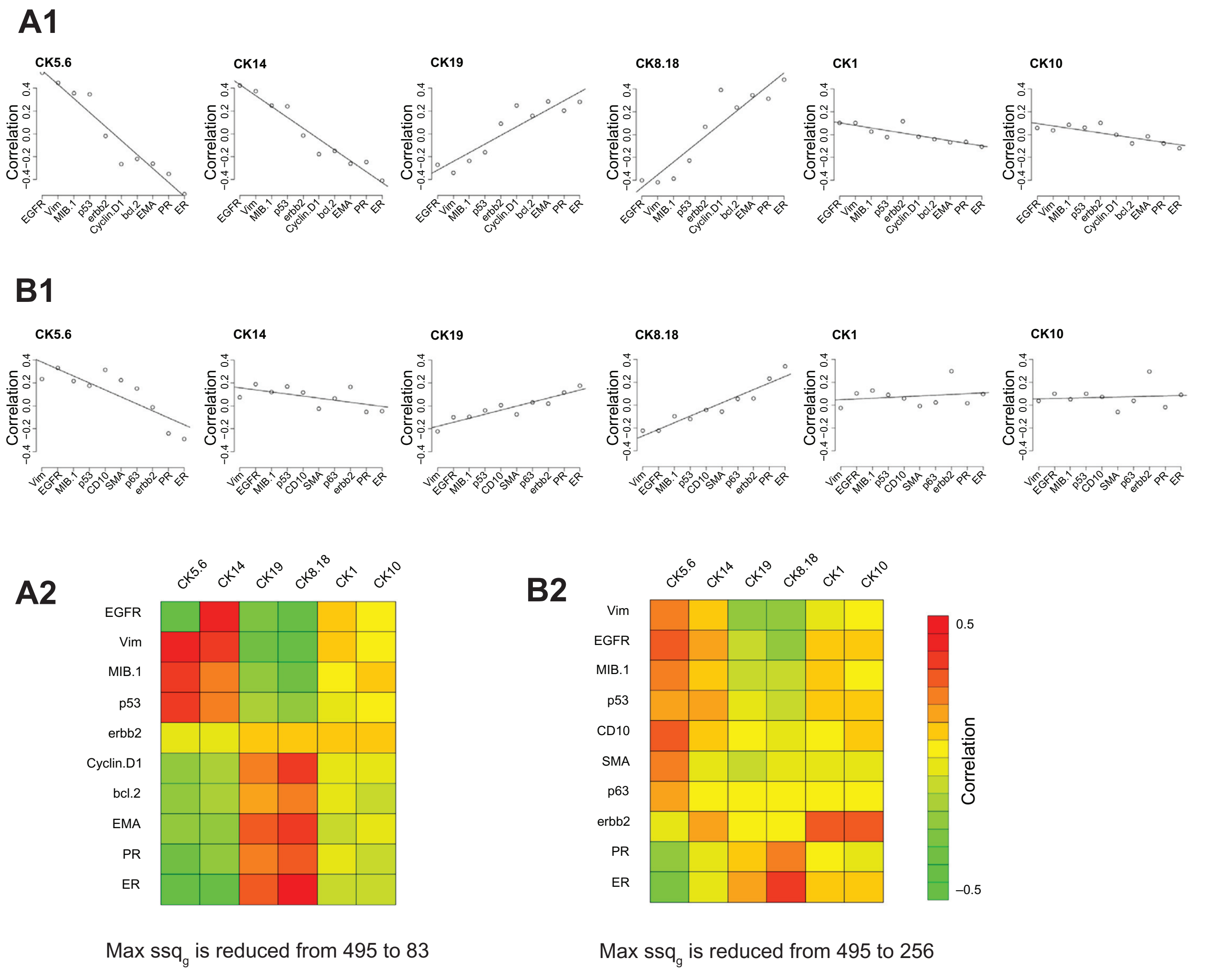

**Figure 3** The optimal caption dependency scheme for the main breast cancer data sets.

**Notes:** The scatter plots on top of the figure are showing the values of the correlation on the *y*-axis according to each cytokeratin reference factor. The test factors are listed on the *x*-axis. For one panel, the rank order of the test factors is always identical (see the Methods section for exemplification). The upper end on the *y*-axis indicates a positive correlation, the lower end an inverse correlation. The order of the ten factors on the *x*-axis depicts the final optimal rank order according to the minimal ssq$_g$ of all regression lines. Cytokeratin (CK) 5/6 and CK8/18 represented the extremes with diametrical opposing regression lines. CK14 and 19 behave similarly compared to CK5/6 and CK8/18, respectively. Reference situations exhibiting a moderate to high slope in the regression line show effects from the optimal rank ordering of the test factors positioned at the extremes of the *x*-axis, while reference situations exhibiting a small slope show that there is no effect at all. (**A1**) M-16; (**B1**) O-16. (**A2**) and (**B2**) are correspondingly showing a visualization of the (ordered) correlation matrices. Further graphs can be seen in supplementary Figure S3.

data matrix might therefore result in smaller (better) or greater (worse) ssq$_g$ solutions. This will test how specific the generated results will be.

A sampling approach without replacement (permutation) was performed on the complete data matrix, while the bootstrap approach (with replacement – approx 37%) was applied separately for each protein staining (see Figure 1, 1B and 3B).

At first, we applied the approach on the level of the raw data. The result for the M-16 collection is very distinct. Performing $10^{+4}$ tests, all the permutation as well as the bootstrap results gave no smaller ssq$_g$ value ($P = 0$, Figure 5A bootstrap, Figure S6A1 and 3). In the case of collection 2 with the seven matching test factors, it behaves differently. A total of 250 of the $10^{+4}$ tests resulted in a smaller ssq$_g$ value ($P = 0.025$, respectively, $P = 0.022$, Figure 5B bootstrap, Figure S6B1 and 3). The combination of collection 1 with collection 2 on the basis of the seven matching test factors (M + O-13) resulted for the sampling approach in five smaller ssq$_g$ values ($P = 5 \times 10^{-4}$).

Performing the approach as described on the level of the correlation data – permutation of the complete matrix or bootstrapping on the reference-factor columns – revealed very similar results. In Figure S6A2 and 4, the results for the M-16 collection are shown. There is no smaller ssq$_g$ like in the raw-data analysis. Looking on collection O-13 (Figure S6B2 and 4), the procedure generates again some smaller ssq$_g$ values ($P = 0.024$, respectively, $P = 0.022$).

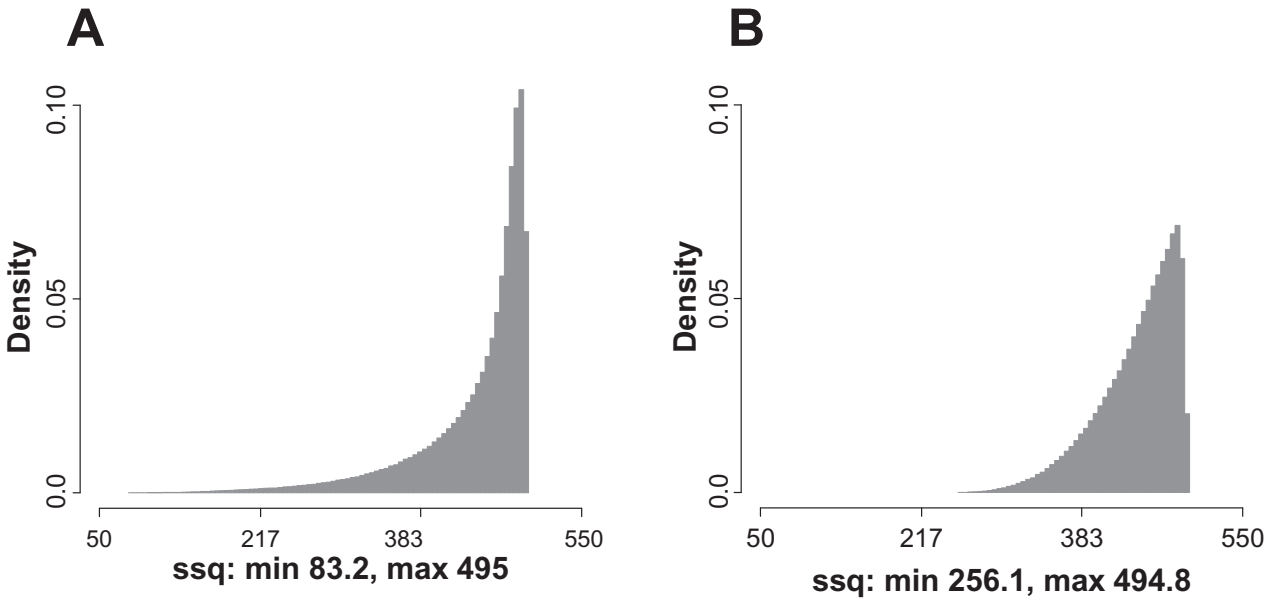

**Figure 4** Basic distributions of the main data sets.
**Notes:** The distributions show all the possible global sum of sum of squares ($ssq_g$) values generated by the permutation procedure. On the *x*-axis, the $ssq_g$ range is shown, while on the *y*-axis the normalized counts are plotted (total number of counts is one). (**A**) M-16; (**B**) O-16. The number of permutations is $3.6 \times 10^6$. The spread of the distributions is different.

Collection M-16, with the CKs acting as reference factors, behaves, concerning the probability values, more ideally than collection O-13 and also the joint collection M + O-13. It makes no difference to apply sampling or bootstrapping procedures on the level of the raw as well on the level of the correlation data. In both levels, the data structure is very sensitive to disorganization. Testing the impact of the sample size on the specificity of the solution, we see that for the given data structure about 200 cases are necessary (Figure 6).

## Analyzing different group compositions

Up to this point, we had chosen a special six of 16 partition of factors, representing our research focus on the CK family. But additional (16 choose 6)-1 = 8007 other partitions are possible to distribute the factors between the test group of size 10 and the reference group of size 6 (Figure 1, 2B). All these combinations represent different views on one and the same physiological situation and can be seen as a low-dimensional projection of the high-dimensional situation.

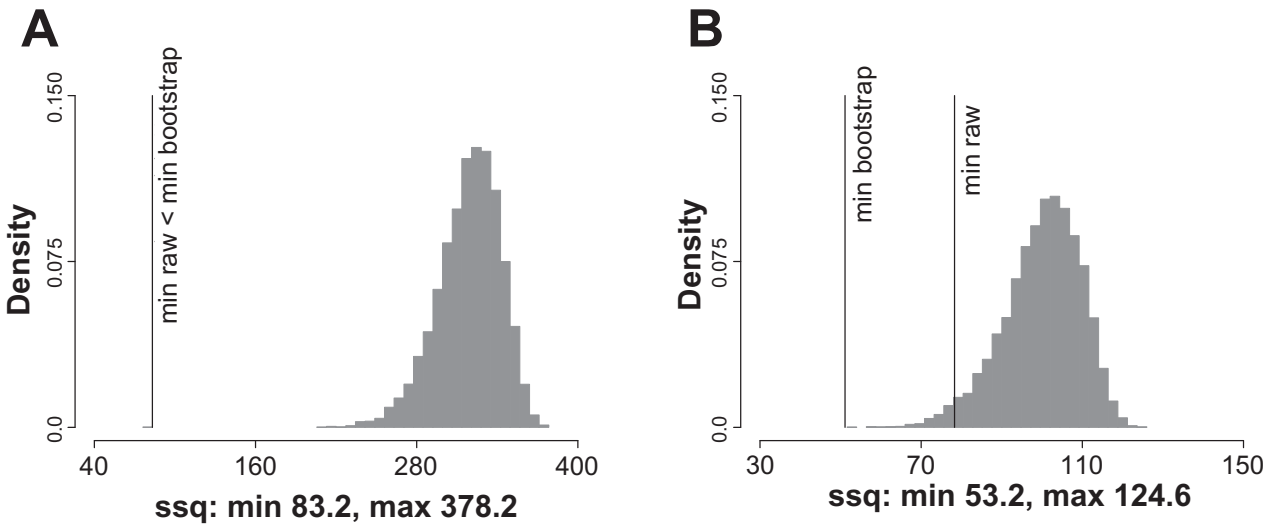

**Figure 5** Distributions of the verification analysis.
**Notes:** The distributions show all the possible global sum of the sum of squares ($ssq_g$) values generated from the respective data sets applying a bootstrap approach to the original or the correlation data (compare Figure 1, 1B and 3B). On the *x*-axis, the $ssq_g$ range is shown, while on the *y*-axis the normalized density is plotted. (**A**) M-16; (**B**) O-16. In both cases, a bootstrap approach per factor on the raw-data matrix is shown. Vertical lines show the position of the minimal $ssq_g$ of the performed bootstrap tests (min bootstrap) and the minimal $ssq_g$ from the analysis of the original unaltered data (min raw).

The question arises as to where our special group partition with its minimal $ssq_g$ is located in the distribution of all possible (six of 16) group partitions. The result based on the data set M-16 (Figure 7A) shows a trimodal distribution. The vertical line denotes the position of our primary combination, which is located near the minimum between the left and the central mode. For the other data sets, this effect is not so clear, indicating an influence of the protein composition and/or the patient collective on the mode structure (Figure S7B–E).

We note that upon testing these results against a permutation/bootstrap sampling of the raw or the correlation matrix, all the multimodal distributions change to unimodal distributions.

At this point, the question arises as to which mode the CK combination of reference factors might belong. The absolutely smallest $ssq_g$ is obviously not linked with the CK reference.

We suspect that the trimodal distribution we are able to observe in collection M-16 represents a superposition of more than one regulatory context. Under this assumption, we should find that our composition should belong to the central mode. To test for this, we constructed all possible replacements of one group member in the primary set of six reference factors, resulting in 60 partitions. We obtained the distribution shown in Figure 7B. The original combination still has the smallest $ssq_g$. Two replacements at a time (675 partitions) resulted in the distribution shown in Figure 7C. The original combination is no longer the lower extreme of the subset in focus, but it is on the lower end of the main cluster of data.

It might be that the trimodal distribution depends on the group size of the test as well as the reference group and not from the underlying data composition. To test for this, we used again collection M-16, but with group sizes ranging from eight of eight to four of twelve (reference to test). Either CK1 and/or 10 were removed or cyclin D1 and/or c-erb-B2 were added. In all situations, the trimodal characteristic remained stable (Figure S8). The maximal $ssq_g$ value shrank from eight of eight to four of twelve, as well as the modes losing their distinct appearance and proportions. This indicates that the mode formation is closely related to the structure of the data set, and not to the partitioning of the data in reference and test set.

These results illustrate two things: (1) the stability of the CK-based six of 16 composition (concerning the smallest $ssq_g$) is limited to one exchange in the reference group, and (2) the modes are composed in a complex way by a superposition of many six of 16 compositions.

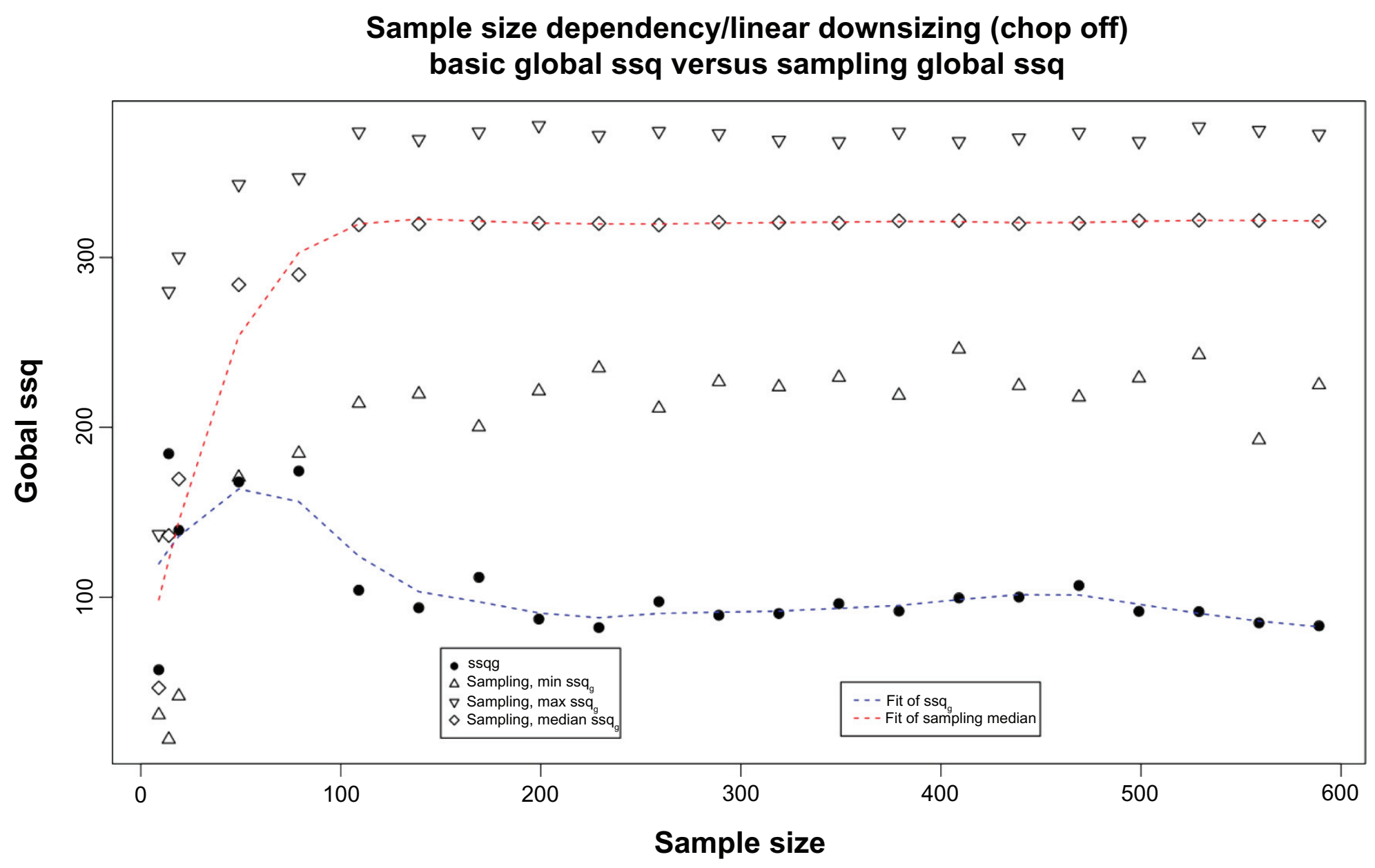

**Figure 6** The impact of the number of samples on the signal-to-noise distance.

**Notes:** The approach compares the bootstrap global sum of the sum of squares ($ssq_g$) values (hollow symbols, red fit line) with best $ssq_g$ value of the given data (solid symbol, blue fit line). Reference and test factors remain unaltered throughout the experiment. In this figure, the M-16 data are presented. A linear downsizing of the full sample size (589) was performed. The plot shows the sample size (*x*-axis) against the respective $ssq_g$ values (*y*-axis). It can be clearly seen that below 200 samples, the signal-to-noise distance decreases, and finally the sampling and the basic results overlap.

# Discussion

The primary focus was put on (1) construction of an optimal rank order of proteins reflecting their cooperative phenomena, and the verification of certain results with the published data, and (2) to give a first draft on the imprinting of the cooperative phenomena in the given data structure. Guided by these objectives, we will discuss our procedure.

## Design of the algorithmic procedure

The optimization algorithm is based on a couple of constraints, in our view being essential to analyze cooperative phenomena. This is not only true for tissue-array data, but more or less for all types of array data. The basic assumption is that a certain protein in a distinct physiological situation has a self-consistent set of relations to a certain number of other macromolecules. This is the basis to construct a rank order reflecting the synergistic, antagonistic, or independent behavior of a protein to other proteins. Secondly, we wanted to correct for noise in the system. This was done by constructing the order by using several joined reference situations. In every situation, the protein must perform as optimally as possible. Therefore, we use the term "optimal" rank order. The local rearrangements of the correlation values to form the rank order illustrate this process. Another reason to use more than one reference situation is to raise the probability that the result is specific for a certain (patho)physiological situation. If a molecular factor has a distinct outcome in every reference situation, the determination is higher and the solution more stable.[26]

The reason we can decipher this optimal rank order by a correlation analysis of one physiological situation is that we have some variation in this state. The variation of 16 factors of one sample is not independent from the variation of 16 factors in another sample. Because of the hidden systemic aspect, all the changes are coordinated. This coordination is imprinting systematic differences in the sample vectors. These differences will be exploited by the optimal rank order to reconstruct dependencies (Figure S1). This in turn means that in an ideal situation (a perfect homogeneous collection of samples), we would fail to detect dependencies with the Pearson correlation.

Generalizing this approach to any other array data type appears readily possible. Naturally, it will be necessary to adapt the procedure to other types of data, sources of data, and possibly different results and interpretations.

Of course, a limitation to this approach is the number of factors, which can be analyzed at once using the computer power available (rate of growth: factorial). We are working on overcoming this limitation by parallelizing the calculation as well as by reducing the combinatorial space. It seems, based on the analysis with three alternative molecular factors, that an assembly process mapping several overlapping analysis approaches together might be feasible.

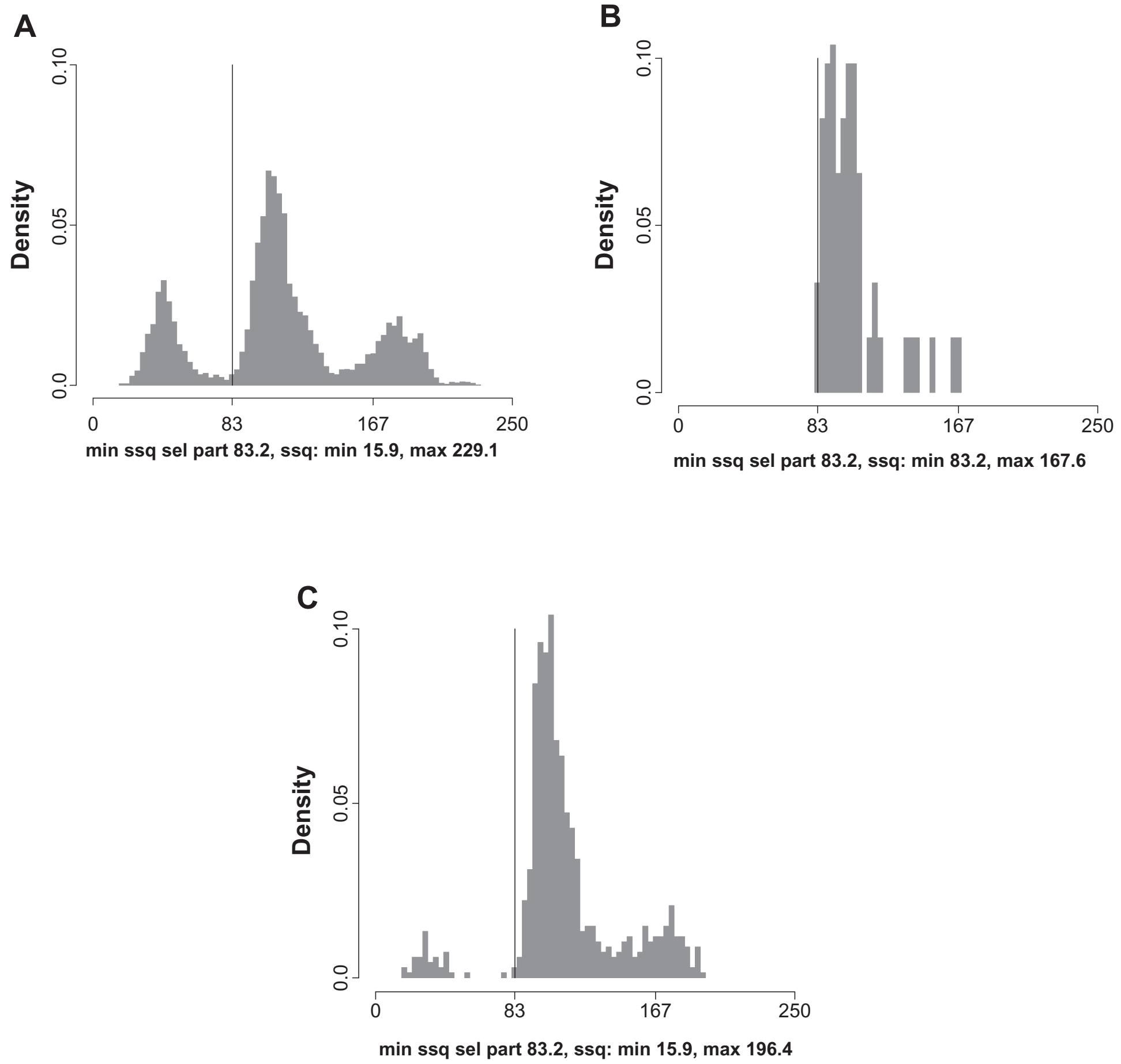

**Figure 7** Distribution of all group combinations and stability of the result against alterations of the groups.
**Notes:** (**A**) All six of 16 combinations of the collective M-16. (**B**) Subset of (**A**) generated by doing an exchange of one group member at once concerning the cytokeratin (CK) six of 16 combination. This means to exchange one CK(CK5/6, CK14, CK8/18, CK19, CK1, and CK10) of the reference group versus one of the test group. The CK combination remains the one with the smallest global sums of the sum of squares ($ssq_g$). (**C**) Subset of (**A**) generated by doing an exchange of two group members at once. The cytokeratin combination is no longer the one with the smallest $ssq_g$. (**A**–**C**) A line denotes the best $ssq_g$ of the six of 16 CK combination.

## Known protein dependencies can be confirmed and supplemented

The adult breast epithelium consists of differentiated luminal glandular and basal myoepithelial cells as well as undifferentiated breast progenitor cells. In the basal compartment, the CK factors CK5, CK14, and smooth-muscle actin are coexpressed, while the CK8/18 are expressed in the mature luminal part. The invasive breast cancer is characterized by the expression of CK5, 14 and absence or the low occurrence of ER/PR/HER2. The mentioned factors can be seen therefore as markers for tumors originating from the respective regions and cellular differentiation stages. The other more weakly characterized factors used in this work might differentiate this picture by supporting one of the extremes or showing no specific participation.

CK5 and CK14, as well as CK8/18 and CK19, show a strong dependence on the optimal rank ordering of dependencies determined between the test and reference factors. This is also once more reconfirmed by a recent study of Shao et al.[27] However, the actual dependence of each pair of factors is diametrically opposite. No obvious trend could be defined for CK1 or CK10. These calculated dependencies are in accordance with protein-expression data described by Abd El-Rehim.[17,28] Chen et al[29] describe this inverse relationship for vimentin, ER, and PR. The presented approach is well suited to extend the analytical approaches shown in, eg, Choo et al[30] and Sarrió et al,[31] because the expression values will be interconnected by an optimal order. The optimal rank order is a definite interpretation of the measured data under a given situation.

For carcinomas with expression of CK5 and CK14, our approach corresponds also on the morphological and clinical level with the published data.[32]

Getting back to the examined collections, identical optimal rank orders could be detected in two completely independent collections of invasive breast cancer cases. We consider the procedure therefore reliable. The small deviations in the slopes of the individual regression lines can be explained by the difference in the sizes of the two invasive breast cancer cohorts, and by their assembly in two different institutes. It is known that institutions usually differ slightly in the classification of borderline cases, leading to different case mixes in the collections used.

Furthermore, the procedure works constitutively. If new factors are added, they are inserted within the existing optimal rank ordering at certain positions, thereby refining this order, but preserving parts or modules of its previous structure.

The insertion of new factors, as shown with the collection of the second institute of pathology, might have an impact on the distribution of the $ssq_g$ of all six of 16 group compositions. There, the mode structure of the distribution might change due to the insertion of factors belonging to a different pathway situation.

Summarizing these observations, the approach is able to analyze the cooperative phenomena of several prominent and some weakly characterized proteins in breast cancer. The calculated results correspond perfectly with already published observations.

## Implications resulting from the initial data type

The tissue microarray data referred to in this experimental setting consist of relative measurements per protein. The dependencies between the proteins are detected on the basis of the imprinted pattern of signal profiles of the observed proteins. This implies that we cannot make assumptions that are linked to absolute concentrations of the metabolites.

## Permutation process and optimization of the rank order

The permutation procedure is used to analyze the complete combinatorial space of protein dependencies to find the most likely dependency order given by the set of array measurements. When differences between a true and a wrong order are small, errors in a measurement may have a profound impact on the resulting rank order. Thus, simple sorting of correlation values might not lead to a true rank order. To correct for this effect, the rank order will be optimized by multiple reference situations. Test and reference set are generated by partitioning the protein measurements into two groups.

The number of reference factors (or reference situations), n, can vary. In the limiting case of one reference factor, we have a trivial ranking, and the correlation value is the only surrogate measure for the dependency. Not the thing we intend. If $n > 1$, we are able to construct in the case of $n + m = \text{even}$: $z = (m + n)/2$ or in the case $n + m = \text{odd}$: $z = (m + n - 1)/2$ group size partitions. The partitions $n > 1$ do correct for errors. If n is growing towards z/2, the $ssq_g$ range will become smaller and the modes will overlap strongly. We have analyzed this effect from $n = 4/m = 12$ to $n = 8/m = 8$ and see an acceptable operating point for the six of 16 partition, taking into consideration the computing time and the biological interest.

The permutation sampling and the bootstrap controls reveal no difference and show the specificity of the process.

If entities broaden up too much, as seen in the collection of the second pathology, the controls will show this by exposing some better $ssq_g$ values. This can be seen as a vital control system proving if the heterogeneity/quality of the presented system is acceptable and whether calculations can be performed.

## Analyzing all the group compositions of the data set

The analysis of all partitions gives a comprehensive insight into the dependency scenario, resulting in a trimodal distribution of the minimal $ssq_g$ values of each analyzed partition. It can be supposed that these three groups in the distribution might reflect three superimposed dependency structures, so the within-group dependency is higher than the between-group dependency. This is underlined by the observation that a randomization of the raw measurements or correlation values will show a change of the observed trimodal distribution to a monomodal distribution. Following this concept, we may find at the lower end of the respective peaks the respective optimal solutions for each individual dependency structure. The dependency between the networks is reflected by the portion of overlap in the distributions tested by the construction of all possible replacements of one group member in the primary set of six reference factors. The stability can be shown for one replacement at a time.

The assumption that this might reflect a superposition of three major dependency structures remains to be justified in a further study with real data and simulated data.

# Conclusion

The algorithm is able to extract an optimal rank order of test factors in a given reference-factor situation. This optimal rank order can be interpreted as a complete net of dependencies between all the analyzed 16 factors. The process needs neither additional information nor any constraints beyond the given immunohistochemical measurements. The definition of an initial grouping is not mandatory. The experimental investigator gets therefore an exciting procedure for their TMA experiments.

For research on invasive breast cancer, this study adds some interesting details on the interaction of the CKs with a couple of other biological factors, and invites the testing of a lot more factors with this combinatorial procedure.

As a side effect, this approach is also able to test if marker panels are relevant and specific for clinical classification purposes. An advanced application of the algorithm in a clinical setting is given in Schymik et al[33] illustrating the versatility of the approach.

## Acknowledgments

We thank Inka Buchroth for the immunohistochemistry. This study was funded in part by the European Commission (LSHCCT-2006-018 814), EuroBoNet, a European Network of Excellence, and in part by the German Federal Ministry of Education and Research BMBF (01GM0869) TranSaRNet.

## Disclosure

The authors report no conflicts of interest in this work.

# Supplementary materials

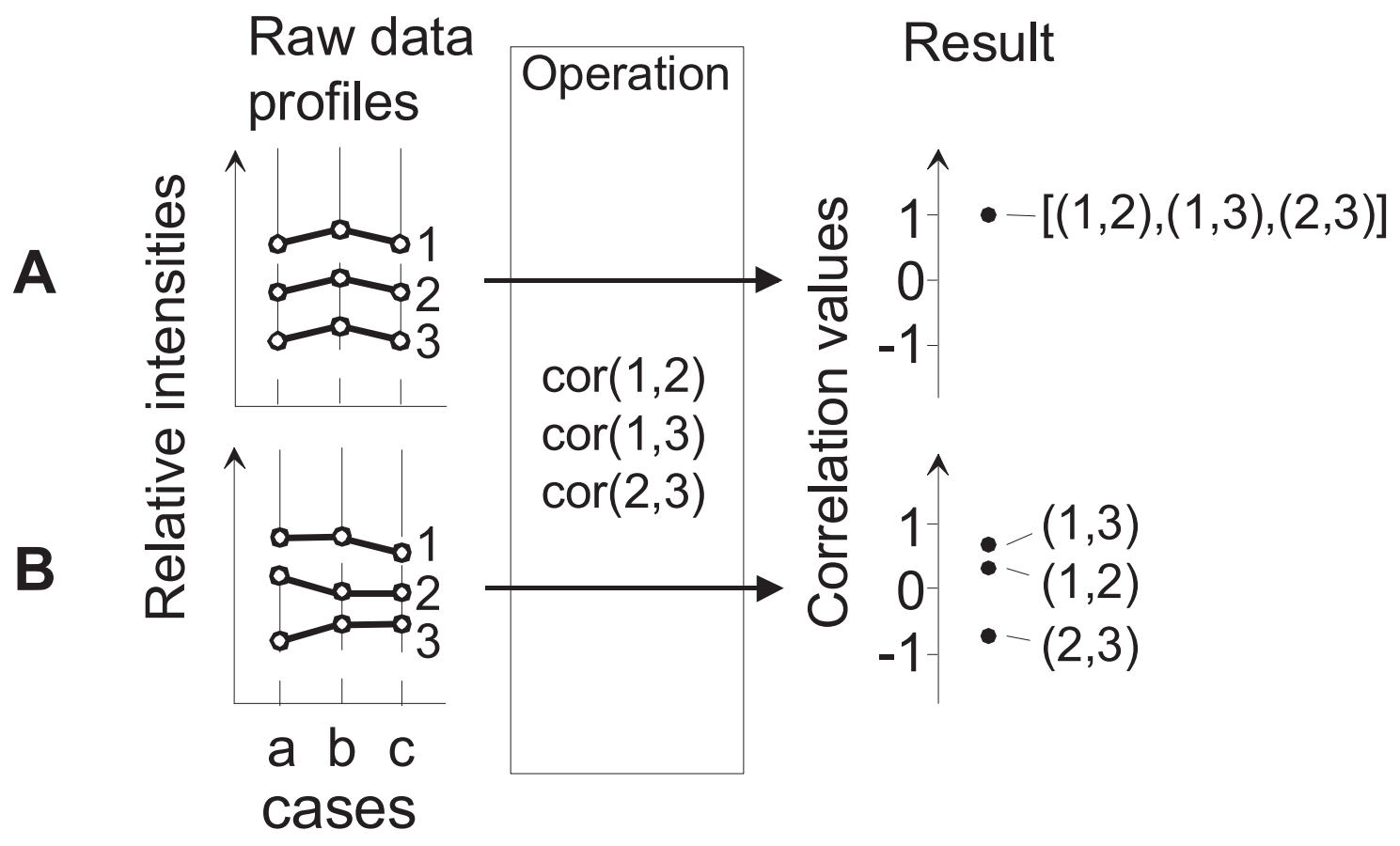

**Figure S1** Consequences in using a certain type of proximity measure.

**Notes:** The selected proximity measure is the correlation according to Pearson.[1] This schematic example shows in panel (**A**) the situations, where all the three biological factors have identical measurement profiles. Only the relative intensity levels differ. The calculation of all possible correlations leads to only one value. In this undifferentiated state we can not distinguish the three data sets. Panel (**B**) shows the case of different profiles for each biological factor. The correlation values for each pair of profiles are different. It is essential for the reliability of the algorithm to have a situation where (a) profiles show variance and (b) the number of cases is great enough to show a specific outcome in the optimization procedure (for this issue cf. Figure 3 and supplementary Figure 2). Note that the Pearson correlation is comparing variances not intensity levels.

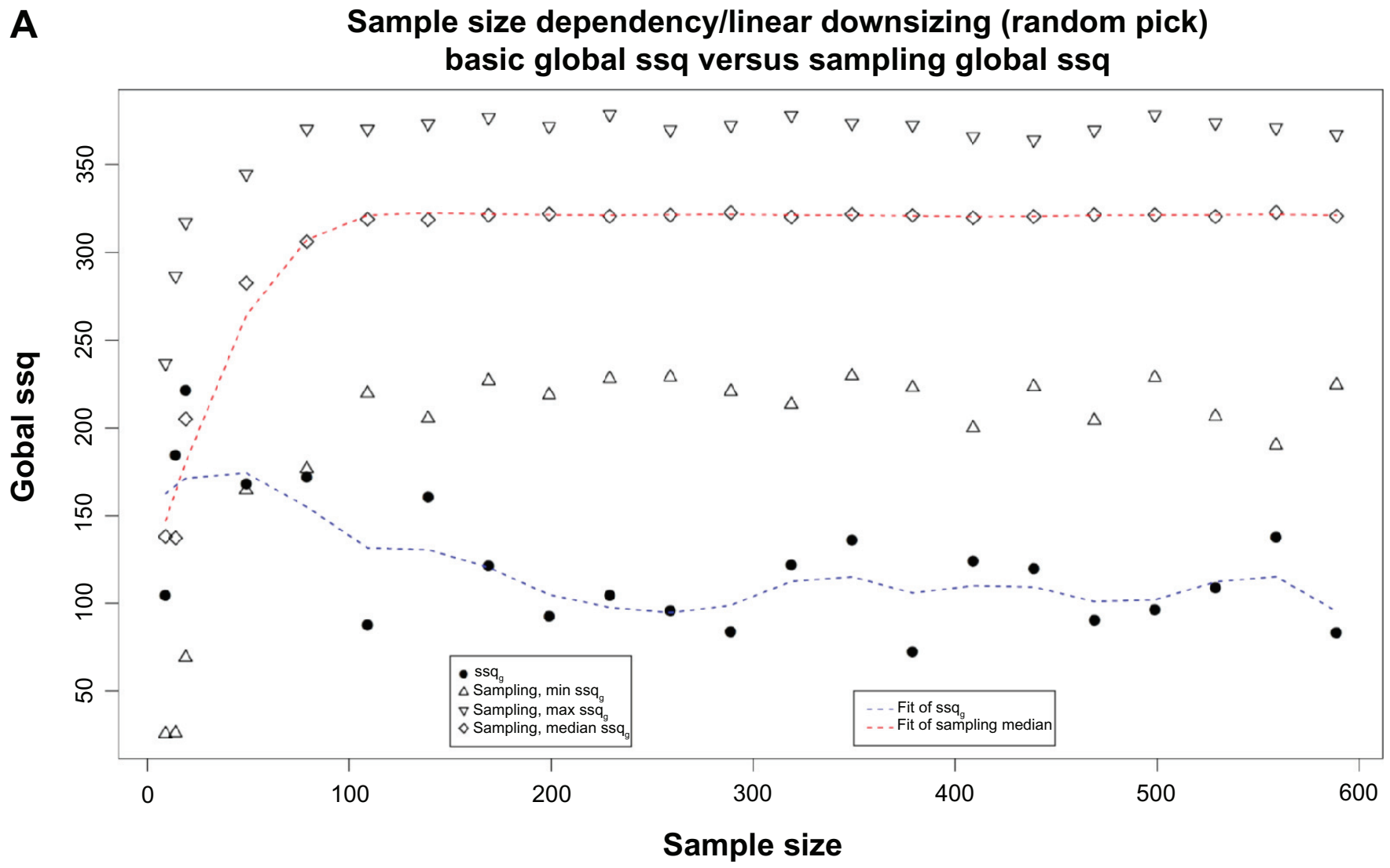

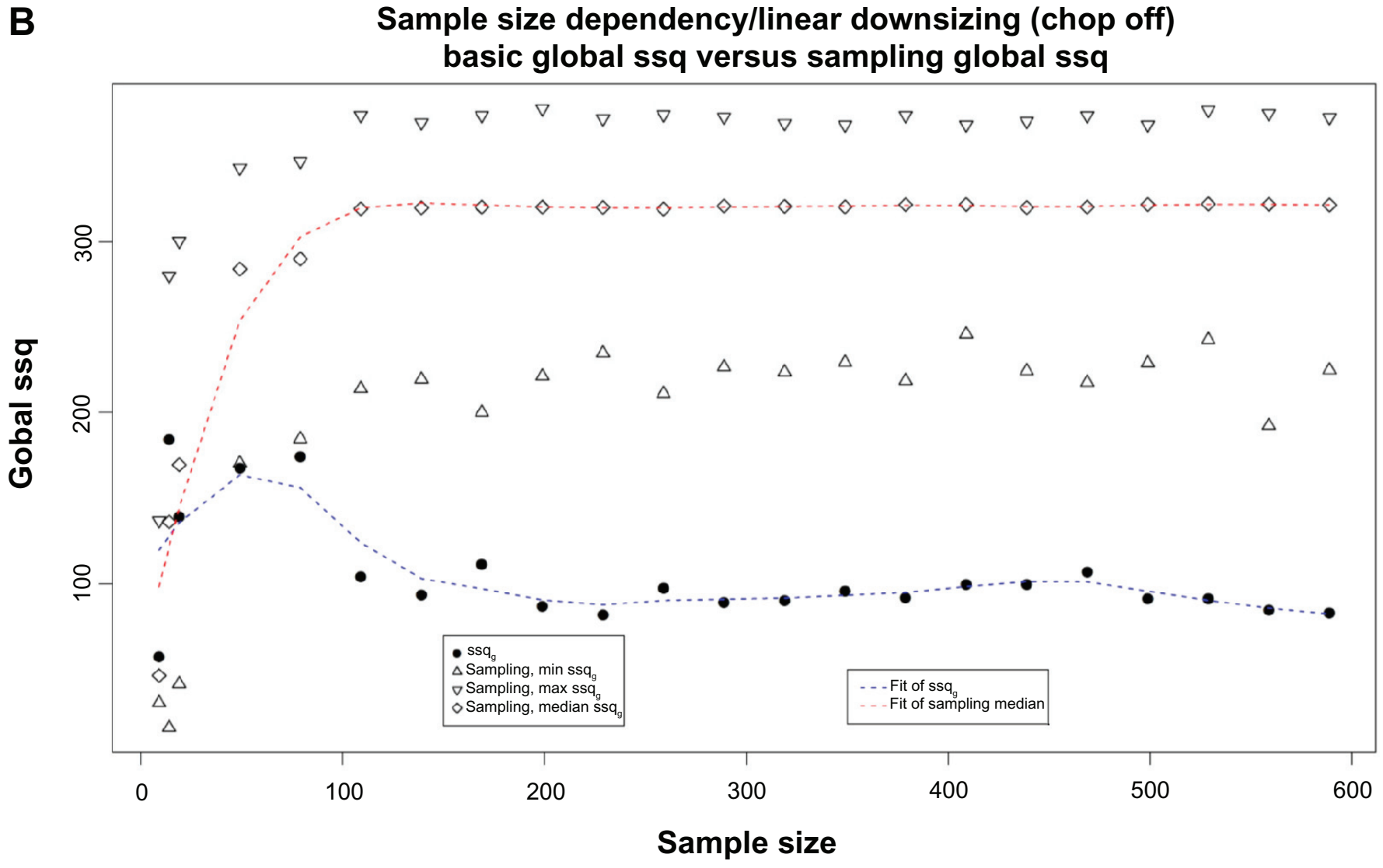

C

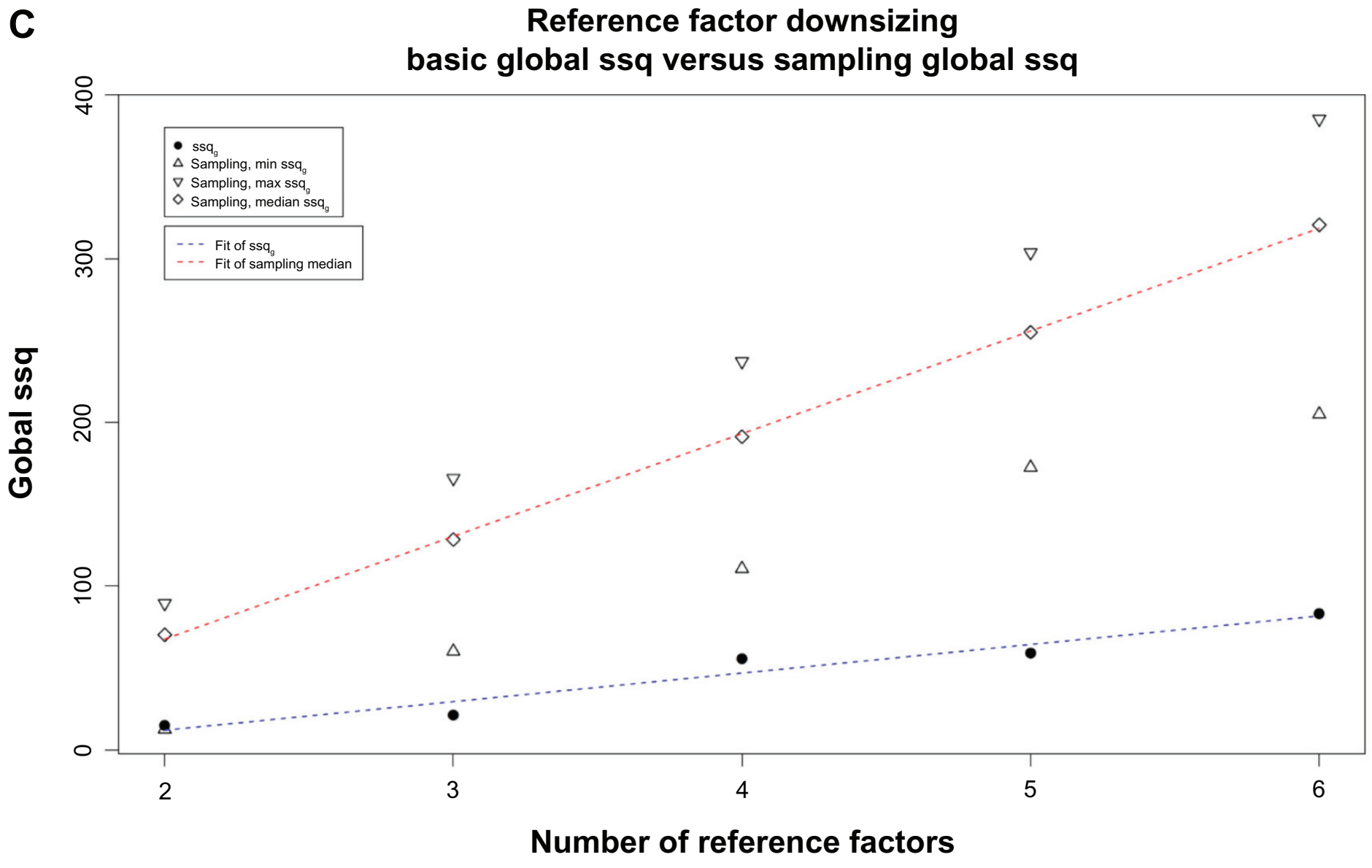

D

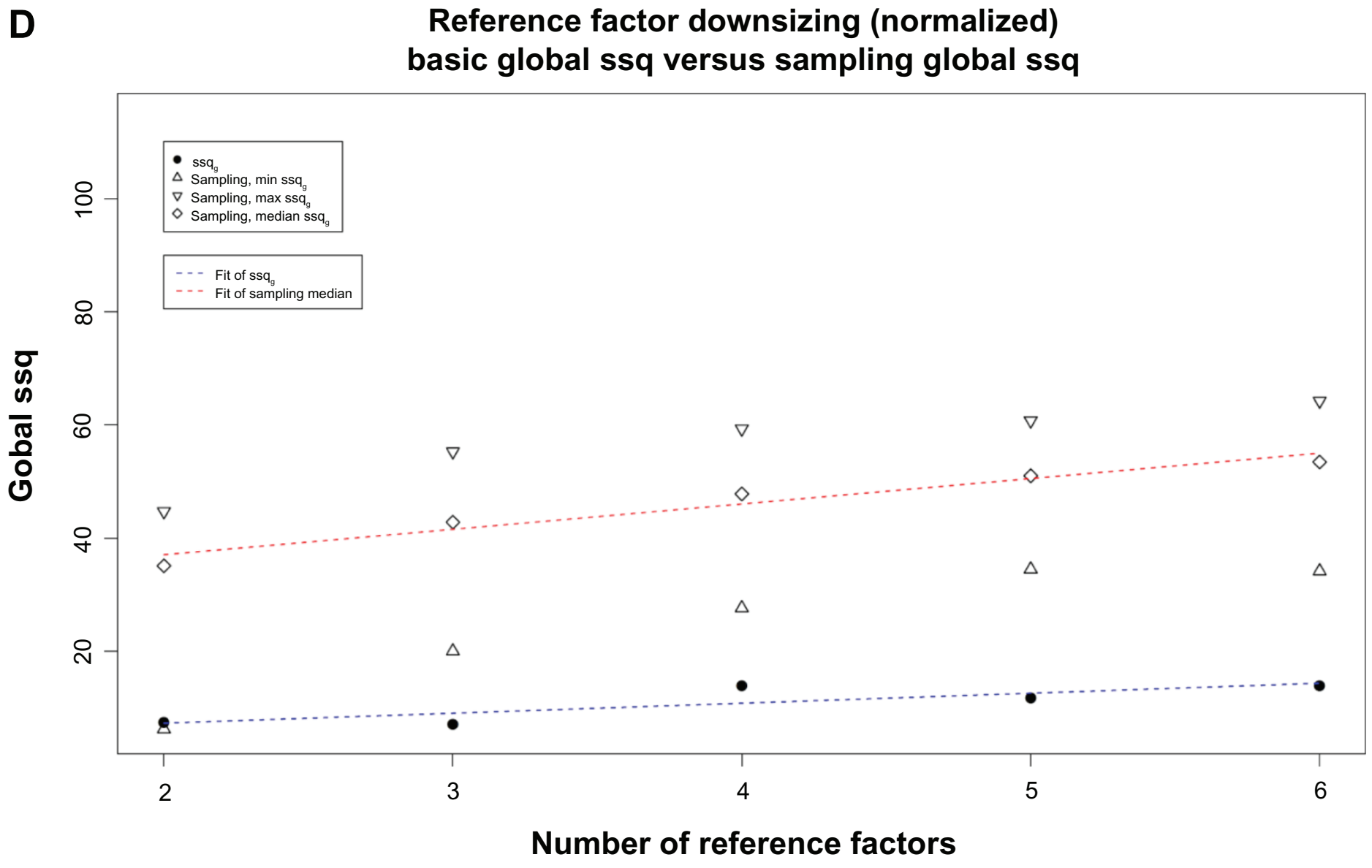

**Figure S2** The impact of the number of samples and reference factors on the signal to noise distance.

**Notes:** Four situations - including the Figure 3 of the main publication - will be presented. (**A**) and (**B**) are illustrating the impact of the sample size on the specificity of the result. The impact is analyzed by applying (**A**) a random sub selections on the full sample size or (**B**) by linearly downsizing the full sample size. (**A**) is therefore showing more fluctuations than (**B**). The signal to noise distance is stable at a sample size of greater 200. (**C** and **D**) are analyzing in a similar way the impact of the number of reference factors. The number of samples is fixed. (**C**) shows the raw analysis while (**D**) shows a normalized version of (**C**). Normalization was performed by division through the number of chosen reference factors. (**A–D**) In all graphs the M-16 data is presented. The bootstrap $ssq_g$ values (hollow symbols, red fit line) will be compared with the best $ssq_g$ value of the given data (solid symbol, blue fit line). (**A, B**) Reference and test factors remain unaltered throughout the experiment. Both plots show the sample size (*x*-axis) against the respective $ssq_g$ values (*y*-axis). (**A**) A linear downsizing of the full sample size (589) was performed. It can be clearly seen, that below 200 samples the signal to noise distance decreases and finally the results overlap. (**B**) Like in (**A**) a downsizing of the full sample size was performed. But at every step the reduced sample was randomly drawn from the full set of 589 samples (‘random pick’, uniform distribution). The variance is more prominent. This is also a good indicator on how stable results are if different compositions of samples are tested. Nevertheless also in this approach, 200 samples remain a good estimate for generating a stable result. (**C** and **D**) The sample size remains constant (589 cases). The reference factors (all cytokeratins) were stepwise reduced from 6 to 2. The test factors remain unaltered. (**C**) shows the raw results while (**D**) shows the normalized version. Both results clearly demonstrate that also in this perspective a higher number of reference factors enlarge the signal to noise distance.

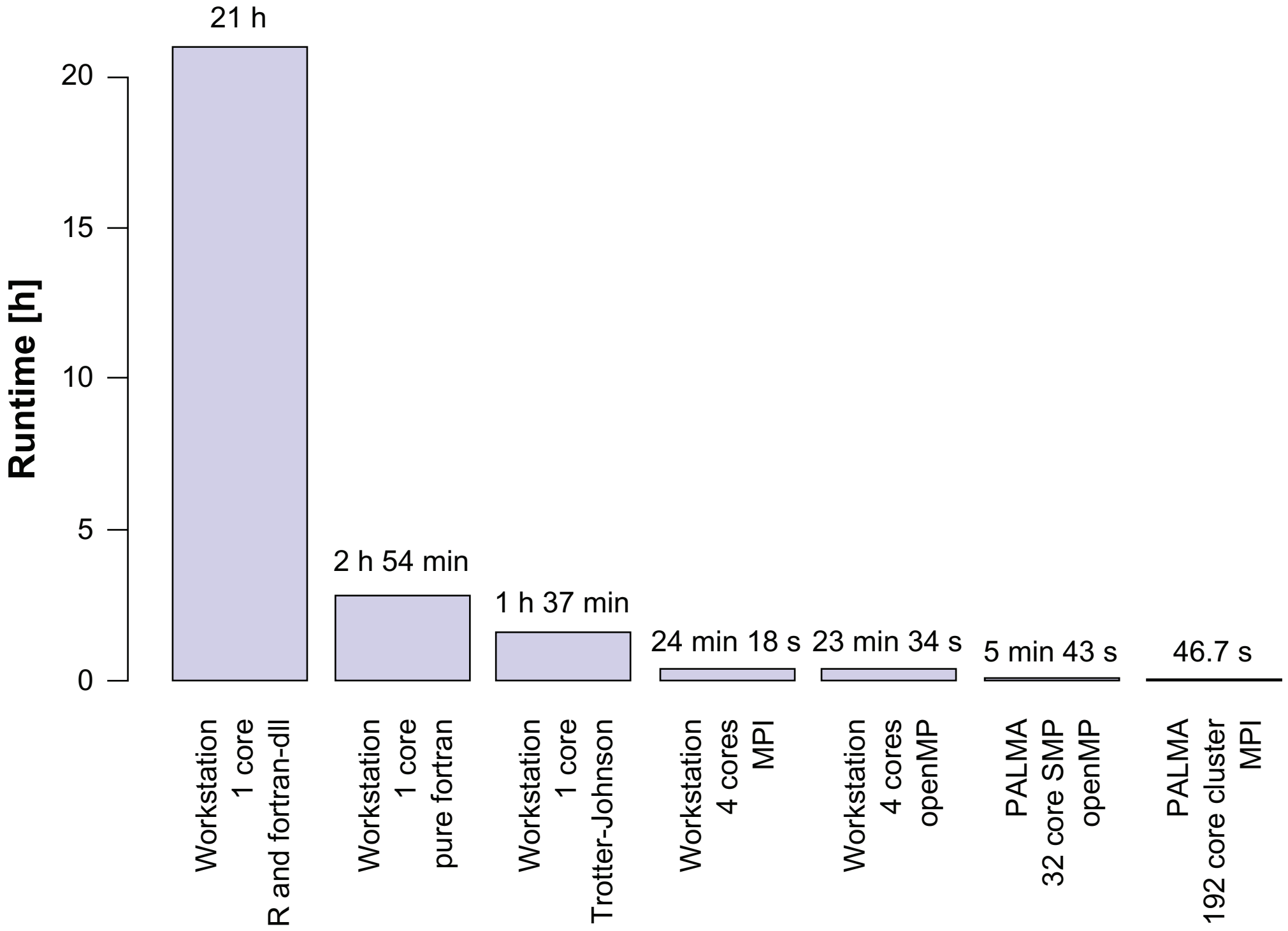

**Figure S3** Computing time constraints of the algorithm.

**Notes:** A runtime comparison based on an example job is shown. 8008 combinations with 6 reference factors out of a total of 16 factors were calculated using some different implementations of the program. The first column from the left represents the implementation, which was used to establish the algorithm and to calculate the presented results. The columns to the right show some ongoing development. The gain is to be up to 1600 times faster compared to the original implementation. But the real challenge is to invest more work in theoretical concepts to select the informative areas of the combinatorial space.

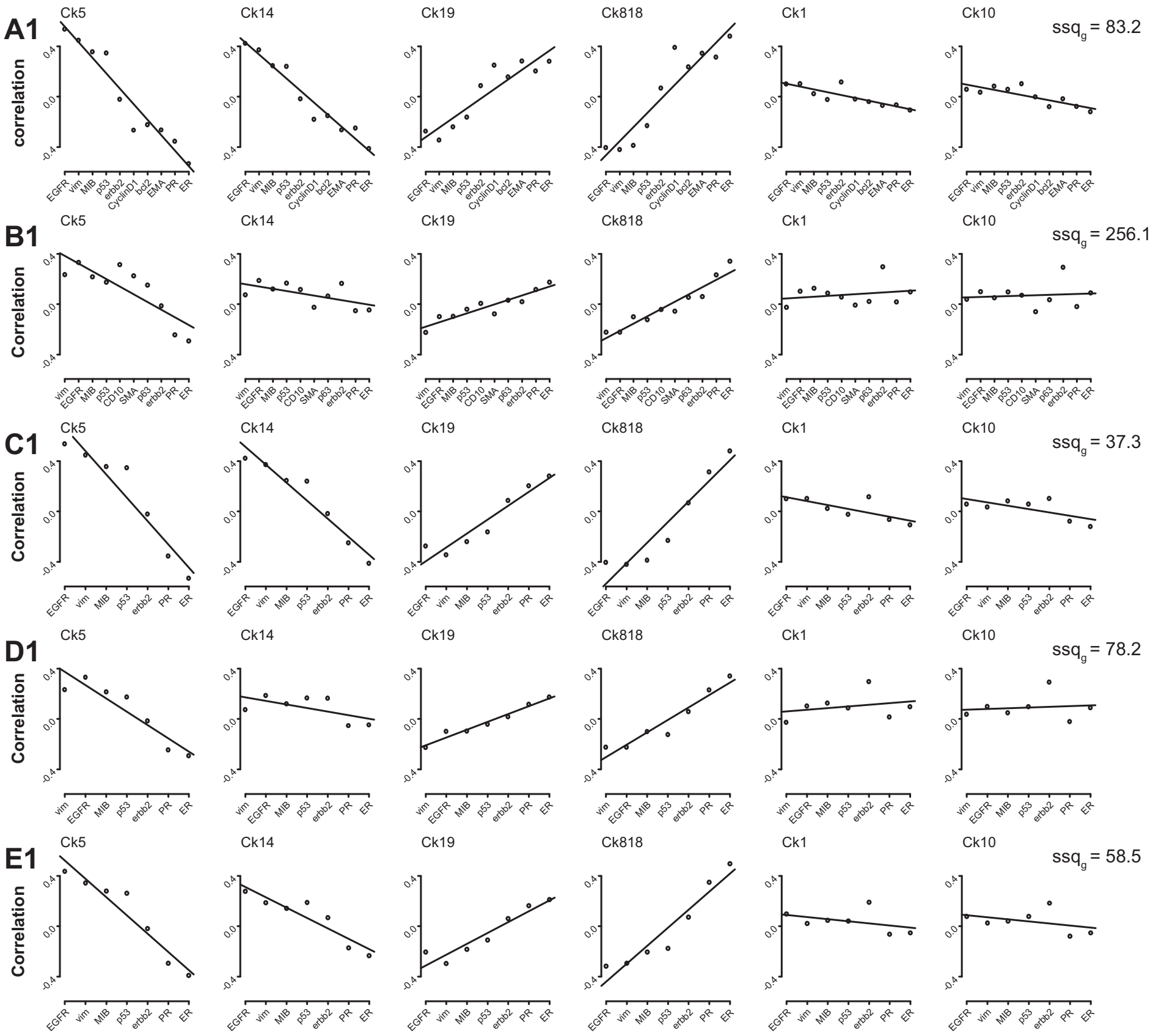

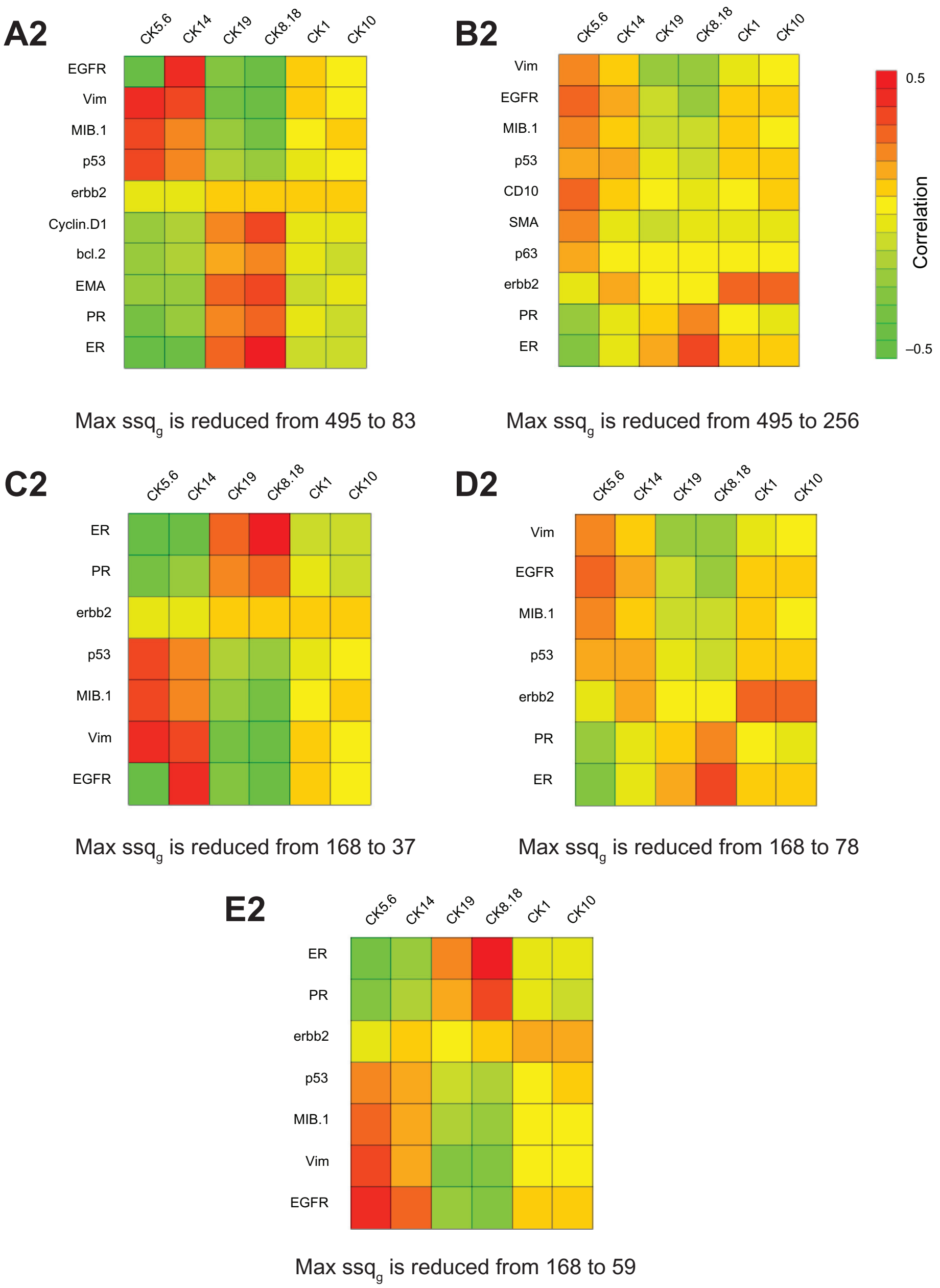

**Figure S4** The optimal dependency scheme for all breast cancer data sets.

**Notes:** This figure corresponds to Figure 3 (and is including Figure 3). The plots are showing the correlation (*y*-axis) according to each reference factor while the test factors are listed on the x axis. For a reference group the order of test factors is always identical (see Methods for exemplification). The order of the test factors on the x axis represents the final optimal rank order. Cytokeratins 5/6 and 8/18 represented the extremes with diametrical opposing regression lines. CK 14 and 19 behave similar compared to CK 5/6 and CK 8/18, respectively. Reference situations exhibiting a moderate to high slope in the regression line show effects from the optimal rank ordering of the test factors positioned at the extremes of the x axis, while reference situations exhibiting a small slope show that there is no effect at all. (**A1**) M 16, (**B1**) O-16, (**C1**) M-13, (**D1**) O-13 and (**E1**) M+O-13. The corresponding graphs on the next page show a visualization of the (optimal ordered) correlation matrices.

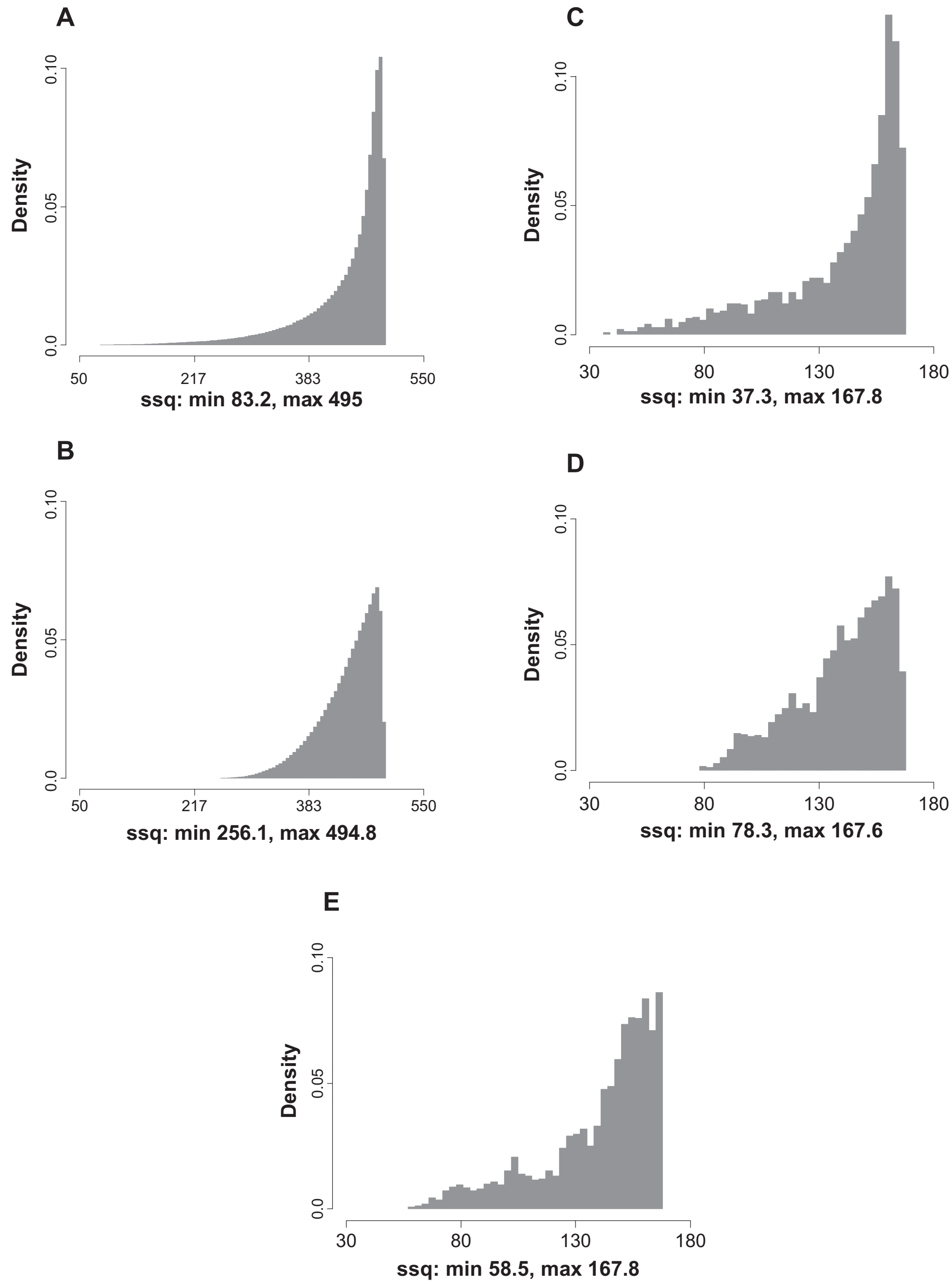

**Figure S5** Basic distributions of all 5 analyzed data sets.

**Notes:** The distributions show all the possible $ssq_g$ generated from the 5 basic data sets. On the x-axis the ssqg range is shown, while on the y-axis the normalized counts are plotted (total number of counts is 1). (**A**) M-16, (**B**) O-16, (**C**) M-13, (**D**) O-13 and (**E**) M+O-13. Depending on the 6 number of factors a different number of permutations is possible: (**A** and **B**) $3.6*10^6$, (**C–E**) 5040. Obviously the fluctuations grow for smaller numbers of permutations.

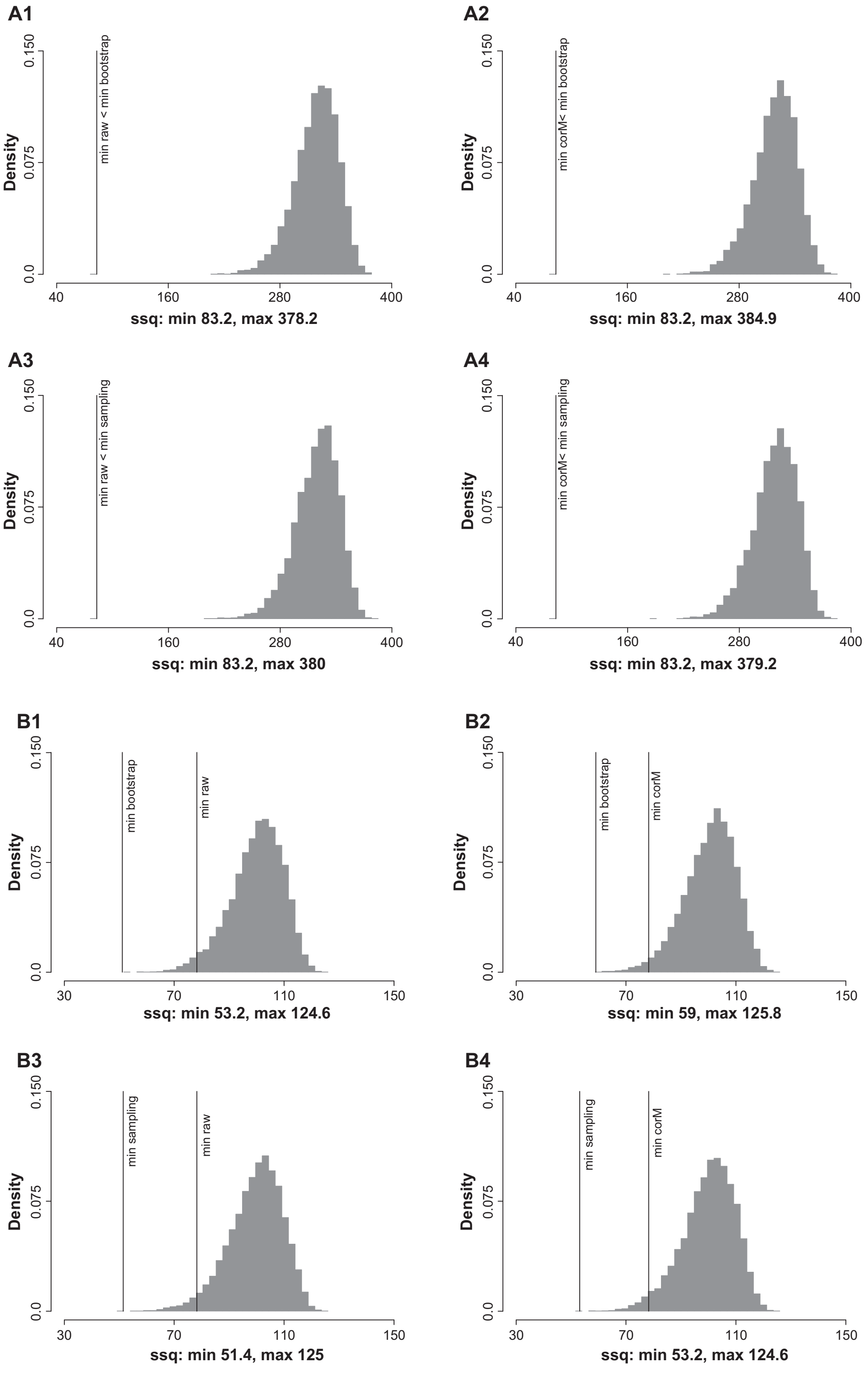

**Figure S6** Distributions of the verification analysis.

**Notes:** The distributions show all the possible $ssq_g$ values generated from the respective data sets applying a bootstrap or a permutation sampling to the original or the correlation data (compare Figure 1-1b,3b). On the x axis the $ssq_g$ range is shown, while on the y axis the counts are plotted. **A1-A4** M-16: (1) bootstrap approach per factor on the raw data, (2) bootstrap approach per reference factor on the correlation data, (3) permutation approach on the complete raw matrix, (4) permutation approach on the complete correlation matrix. **B1-B4** O-13: (1) bootstrap approach per factor on the raw data, (2) bootstrap approach per reference factor on the correlation data, (3) permutation approach on the complete raw matrix, (4) permutation approach on the complete correlation matrix. Vertical lines show the position of the minimal $ssq_g$ of the performed sampling tests (min. bootstrap or min. permutation sampling), and the minimal $ssq_g$ from the analysis of the unaltered data (min. raw) respectively (min. corM) for sampling on the level of the correlation matrix.

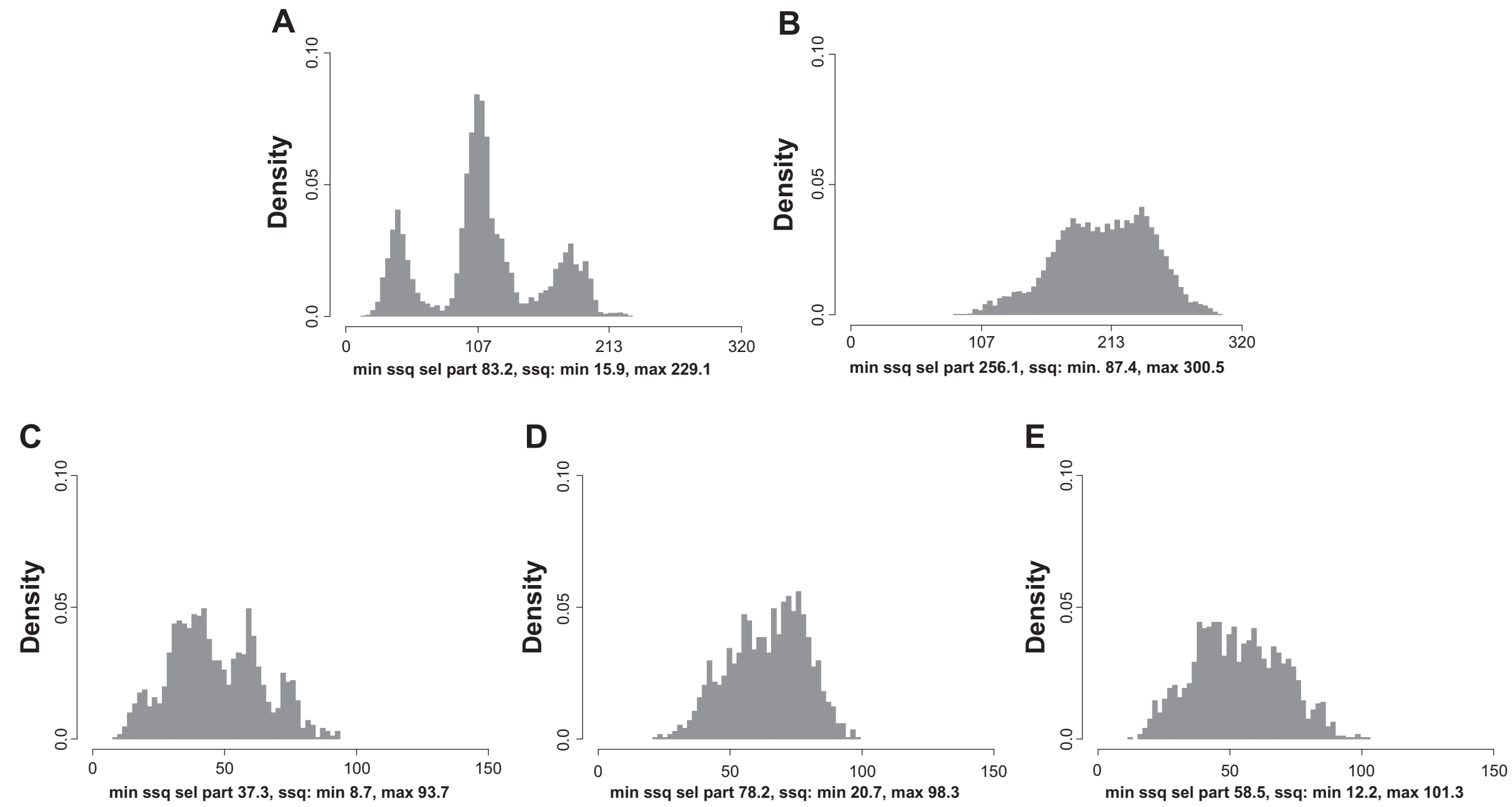

**Figure S7** Distribution of all combinations for all 5 collections.

**Notes:** On the *x*-axis the $ssq_g$ range is shown, while on the *y*-axis the counts are plotted. (**A**) M-16, (**B**) O-16, (**C**) M-13, (**D**) O-13 and (**E**) M+O-13. 'min. ssq sel. part.' means: Minimal $ssq_g$ of the a priory chosen cytokeratin combination. 'ssq min.' respective 'max.' denote the $ssq_g$ spread of the distribution.

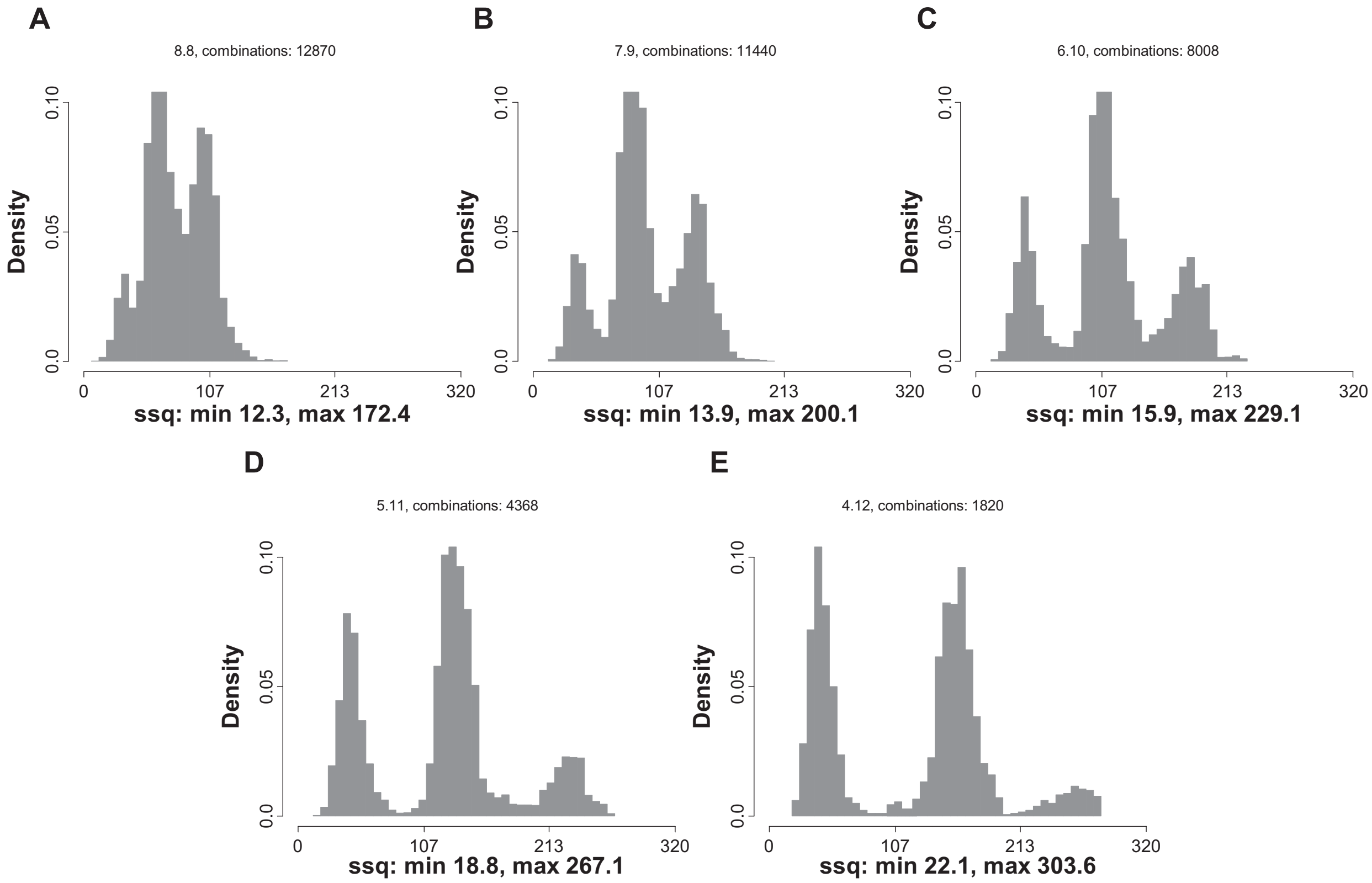

**Figure S8** Testing the variability of the trimodal distribution by altering the size of the reference and test set.

**Notes:** On the *x*-axis the $ssq_g$ range is shown, while on the *y*-axis the counts are plotted. The analysis is based on collection M-16. The group size was varied from (**A**) 8 reference/8 test members to (**E**) 4 reference/12 test members. 'ssq min.' respective 'max.' denote the $ssq_g$ spread of the distribution. On top the number of combinations building the distribution is noted.

**Table S1** Pretreatment conditions, source, and dilution of primary antibodies used for the immunohistochemical staining

| Antibody | Source | Clone | Pretreatment (min) | Dilution | C |
|---|---|---|---|---|---|
| CK1 | Novocastra | 34ßB4 | S (30) Citrate buffer | 1:150 | 1 + 2 |
| CK5 | Dako | D5/16B4 | S (10) Citrate buffer | 1:80 | 1 + 2 |
| CK8/18 | Dianova | 5D3 | S (10) Citrate buffer | 1:40 | 1 + 2 |
| CK10 | Dako | DE-K10 | S (30) Citrate buffer | 1:400 | 1 + 2 |
| CK14 | Dianova | LL002 | S (30) Citrate buffer | 1:50 | 1 + 2 |
| CK19 | Quartett | KS19.1 | S (30) Citrate buffer | 1:80 | 1 + 2 |
| ER | Novocastra | 6F11 | S (10) | 1:30 | 1 + 2 |
| PR | Novocastra | 16 | S (10) | 1:200 | 1 + 2 |
| Ki-67 | Dako | MIB-1 | S (10) Citrate buffer | 1:40 | 1 + 2 |
| c-erb-B2 | Dako | [polyclonal] | S (30) Citrate buffer | 1:400 | 1 + 2 |
| EGFR | Ventana | 3C6 | S (10) Citrate buffer | 1:100 | 1 + 2 |
| Vimentin | Ventana | V9 | S (30) Citrate buffer | 1:1000 | 1 + 2 |
| SMA | Dako | A14 | S (30) Citrate buffer | 1:4000 | 2 |
| CD10 | Novocastra | 56C6 | S (30) Citrate buffer | 1:40 | 2 |
| p53 | Dako | DO-7 | S (10) Citrate buffer | 1:500 | 1 + 2 |
| p63 | Dako | 4A4 | S (30) Citrate buffer | 1:100 | 2 |
| BCL2 | Dako | 124 | S (30) Citrate buffer | 1:500 | 1 |
| Cyclin D1 | NeoMarkers | SP4 | S (30) Citrate buffer | 1:25 | 1 |
| EMA | Dako | E29 | S (30) Citrate buffer | 1:2000 | 1 |

**Notes:** "C" denotes if the antibody was used with collection 1 ("1") or with collection 2 ("2") or in both collections ("1 + 2"). In the Pretreatment column, S means steamer.
**Abbreviations:** ER, estrogen receptor; PR, progesterone receptor; SMA, smooth-muscle actin; EMA, epithelial membrane antigen.